\documentclass[prd,aps,a4paper,floatfix,amsmath,amssymb,twocolumn]{revtex4-2}
\usepackage{mathtools}
\usepackage[margin=3cm]{geometry}
\usepackage{graphicx}
\usepackage[pdftex,dvipsnames]{xcolor}
\usepackage[pdftex,breaklinks,colorlinks,
linkcolor=Blue,
citecolor=teal,
anchorcolor=red,
urlcolor=cyan]{hyperref}

\begin{document}
	

\title{Fully constrained, high-resolution shock-capturing,
	formulation of the Einstein-fluid equations in $2+1$ dimensions}

\author{Carsten Gundlach, Patrick Bourg and Alex Davey}
\affiliation{Mathematical Sciences, University of Southampton,
	Southampton SO17 1BJ, United Kingdom}
\date{24 June, revised 07 August, 2021}

\begin{abstract}
	
Four components of the axisymmetric Einstein equations in
$2+1$ dimensions with negative cosmological constant can be written as
$\nabla_aM=\dots$ and $\nabla_aJ=\dots$, where the dots stand for
stress-energy terms, and $M$ and $J$ are scalars. In vacuum, they
reduce to the constant mass and angular momentum parameters of the BTZ
solution of the same name. The integrability conditions for the
Einstein equations give rise to two conserved stress-energy currents
$\nabla_aj^a_{(M)}=0$ and $\nabla_aj^a_{(J)}=0$. The angular momentum
current is just the Noether current due to axisymmetry, but the mass
current is unexpected in the presence of rotation. The conserved
quantity $M$ exists in all dimensions in spherical symmetry, known as
the Misner-Sharp, Hawking or Kodama mass, but in $2+1$ dimensions $M$
exists also in axisymmetry, even with rotation. We use $M$ and $J$ to
give a fully constrained formulation of the axisymmetric Einstein
equations in $2+1$ dimensions, where the Einstein equations are solved
by explicit integration from the center along time slices. We use the
two conserved matter currents in the construction of a high-resolution
shock-capturing formulation of the Einstein-perfect fluid system, in
which $M$ and $J$ momentum are then exactly conserved by construction. We
demonstrate convergence of the code in the test cases of generic
dispersion and collapse and stable and unstable rotating stars.
	
\end{abstract}

\maketitle

\tableofcontents

\section{Introduction}

We present a formulation of the Einstein equations with matter
and a negative cosmological constant in $2+1$ dimensions, restricted to
axisymmetry, that is fully constrained, in the sense that the Einstein
equations can be solved by explicit radial integration along time
slices to find the metric on that time slice. 

We also present a numerical implementation of this formulation where
the matter is a perfect fluid with the linear (ultrarelativistic)
 equation of state $P=\kappa\rho$. We demonstrate convergence of this
scheme in a number of test cases with $\kappa=1/2$: rotating collapse,
rotating strong field noncollapse, and the time evolution of both
stable and unstable rotating stars, perturbed slightly.

In a companion paper, we shall use this code to investigate critical
phenomena at the threshold of prompt collapse in this system. 

Our numerical implementation could be generalized straightforwardly to
any barotropic or hot perfect fluid equation of state, and our
numerical implementation of the Einstein equations to any other
matter. 

As the starting point for our formulation, we carry out a reduction of
the covariant Einstein equations under the axisymmetry, with barred
quantities referring here and later to the reduced 2-dimensional
spacetime. In axisymmetry in $2+1$ dimensions, there are six
independent components of the Einstein equations. Four of these can be
written as $\bar\nabla_aM=\bar\epsilon_{ab}\bar j^b_{(M)}$ and
$\bar\nabla_aJ=\bar\epsilon_{ab}\bar j^b_{(J)}$, where
$\bar\epsilon_{ab}$ is the 2-dimensional volume form. The left-hand
sides are defined in terms of the Killing vector $\xi^a$ of
axisymmetry and the ``area radius'' $R$ defined by the length of the
closed symmetry orbits. The right-hand sides are the contraction of
the stress-energy tensor with two vectors also made from $\xi^a$ and
$R$.

This tells us two things: $\bar j^a_{(M)}$ and $\bar j^a_{(J)}$ are
conserved matter currents, and $M$ and $J$ are nontrivial quasilocal
(local in the reduced spacetime)
metric invariants that are constant in vacuum. (They
reduce to the constant mass and angular momentum parameters of
the same name in the B\~anados-Teitelboim-Zanelli (BTZ family of
axistationary metrics \cite{Banados92}.)

To stress how unexpected this rich geometrical structure of
axisymmetry in $2+1$ dimensions is, we remind the reader what parts of
it are known in other situations. The local mass $M$ exists, and is
linked to a conserved matter current $j^a_{(M)}$, in spherical
symmetry in any dimension, and is then known as the Kodama \cite{Kodama80}
or generalized Misner-Sharp \cite{Maeda08} mass. The
current arises as the contraction of the stress-energy tensor with a
certain vector field, but this is not a Killing vector field. 
The conserved angular momentum matter current $j^a_{(J)}=T^{ab}\xi_b$
exists in axisymmetry, also in any dimension. However, the local angular
momentum $J$ exists only in $2+1$ dimensions. Moreover, in $2+1$
dimensions only, $M$ and its current exist in axisymmetry even with
rotation.

The structure of the paper is as follows. In Sec.~\ref{section:section2}
we derive the quantities $M$ and $J$ and their
underlying currents in the reduction approach. We use these two
conservation laws, plus a balance law for radial momentum, to
formulate the fluid evolution equations. (To generalize from a
barotropic to a hot equation of state, we would only need to add the rest
mass conservation law.) 

In Sec.~\ref{section:section3} we then introduce specific coordinates on
the reduced spacetime, namely a radial coordinate $r$ linked in a
fixed way to the area radius $R$, and a time coordinate $t$ that is
normal to $R$ (polar time slices). The full metric on a time slice can
then be obtained from suitable fluid variables on that slice by
integration over $r$ (starting from a
regular center). In this form, the Einstein equations look quite
similar to those in polar-radial coordinates in spherical symmetry (in
any dimension).

Section~\ref{section:section4} describes our numerical implementation in
detail. In particular, we discretize the integration of the currents
to obtain $M$ and $J$ so that the latter are conserved exactly. This
is particularly important where $M\simeq 0$ but its sign matters
because black holes can form only for $M>0$ (we use the BTZ convention
where its value in vacuum adS3 is $-1$).
Similarly, for rapidly rotating collapse it will matter if $J$
is larger or smaller than $M$. For the fluid evolution, we use an
evolve-reconstruct-limit approach with a simple approximate Riemann
solver. In several details, we follow methods of \cite{Neilsen00} for
ultrarelativistic fluid collapse in spherical symmetry in 3+1
dimensions.

Section~\ref{section:section5} describes numerical tests. To allow black
holes to exist in $2+1$ dimensions, we assume a negative cosmological
constant throughout. We show that, at least for sufficiently short
times and away from the numerical outer boundary, all variables
converge pointwise to second order. In some situations, the
rate of convergence goes down to first order after numerical error
from our ``copy'' numerical outer boundary condition dominates the
error budget. We use five different tests: generic rotating initial
data that disperse and collapse respectively, and slightly perturbed
stable and unstable rotating stars, the latter perturbed so that they
either collapse or begin highly nonlinear oscillations. All
regular axistationary solutions with finite $M$ and $J$ (``rotating
stars'') in $2+1$ dimensions with negative cosmological constants, and for
arbitrary barotropic equation of state, and $P=\kappa\rho$ in
particular, were classified in \cite{Carsten20}, building on
earlier work in \cite{Cataldo04}. Here we give numerical evidence
for a conjecture made there, that where there are two stars with the
same $M$ and $J$, the more compact one is unstable and the less
compact one stable.

Section~\ref{section:section6} contains our conclusions.

\section{Geometric description of the model}
\label{section:section2}

\subsection{Axisymmetry in $2+1$ spacetime dimensions}

We consider axisymmetric solutions of the Einstein equations in $2+1$
dimensions with negative cosmological constant $\Lambda=:-1/\ell^2$, 
\begin{equation}
G_{ab} + \Lambda g_{ab}=8\pi T_{ab}.
\end{equation}
We set $c=G=1$ throughout.
Let $\xi^a$ be the Killing vector defining the axisymmetry.
Its length defines the area radius
\begin{equation}
\xi^a\xi_a=:R^2
\end{equation}
as a scalar. We define a local
angular momentum $J$ geometrically as the twist (a scalar in $2+1$
dimensions) of the Killing vector,
\begin{equation}
J:=\epsilon^{abc}\xi_a\nabla_b\xi_c,
\end{equation}
where $\epsilon_{abc}$ is the volume form implied by the metric
$g_{ab}$. We define a local mass function $M$ in terms of $J$
and $R$ as
\begin{equation}
M:={R^2\over\ell^2}+{J^2\over 4 R^2}-(\nabla_aR)(\nabla^aR).
\end{equation}

Following Geroch, we define the metric in the reduced $1+1$-dimensional
spacetime of orbits
\begin{equation}
\bar g_{ab}:=g_{ab}-R^{-2}\xi_a\xi_b,
\end{equation}
so that $\bar g_{ab}\xi^b=0$,
the corresponding volume form
\begin{equation}
\bar \epsilon_{ab}:=R^{-1}\epsilon_{abc}\xi^c,
\end{equation}
and the corresponding covariant derivative operator $\bar\nabla_a$ by
\begin{equation}
\bar{\nabla}_a :=\perp \nabla_a \perp
\end{equation}
where $\perp$ stands for contraction with $\bar g_a{}^b$ on all indices.

Four linear combinations of components of the Einstein
equations can then be written as
\begin{eqnarray}
\label{DaJ}
\bar{\nabla}_a J&=&-16\pi R\, \bar\epsilon_{ab}\, j_{(Z)}^b, \\
\label{DaM}
\bar{\nabla}_a M&=&-16\pi R\, \bar\epsilon_{ab}\, j_{(\Omega)}^b
\end{eqnarray}
Clearly the currents $j_{(Z)}^a$ and $j_{(\Omega)}^a$ are conserved in the
sense that
\begin{eqnarray}
\bar{\nabla}_a (R\,j^a_{(Z)})&=0, \\
\bar{\nabla}_a (R\,j^a_{(\Omega)})&=0, 
\end{eqnarray}
or equivalently
\begin{eqnarray}
\label{DajaZ}
\nabla_aj^a_{(Z)}&=&0, \\
\label{DajaOmega}
\nabla_aj^a_{(\Omega)}&=&0.
\end{eqnarray}

The angular momentum and mass currents introduced above are given by
\begin{eqnarray}
j_{(Z)}^b&:=&V_{(Z)a}T^{ab}, \\
j_{(\Omega)}^b&:=&V_{(\Omega)a}T^{ab},
\end{eqnarray}
where
\begin{equation}
V_{(Z)}^a:=\xi^a
\end{equation}
and 
\begin{equation}
V_{(\Omega)}^a:=V_{(X)}^a+{J\over 2R^2}V_{(Z)}^a,
\end{equation}
with 
\begin{equation}
V_{(X)}^a:=\bar\epsilon^{ab}\nabla_b R, 
\end{equation}
or equivalently 
\begin{equation}
V_{(\Omega)}^a=R^{-2}\left(\epsilon^{abc}\xi^d+{1 \over 2}
	\xi^a\epsilon^{bcd}\right) \xi_b \nabla_c \xi_d.
\label{VOmega}
\end{equation}

The conservation law (\ref{DajaZ}) follows directly from the fact that
$V_{(Z)}^a:=\xi^a$ is a Killing vector, but (\ref{DajaOmega}) is less
obvious. In spherical symmetry, $j^a_{(\Omega)}$ and $M$ are known
generalizations of the Kodama conserved current and mass
\cite{Kodama80} from 3+1 to arbitrary dimensions
\cite{Maeda08}.

While this paper was under review, a paper has appeared
\cite{Kinoshita21} that independently identifies the same generalised
Kodama vector. It is given there in the form
\begin{equation}
V^a_{(\Omega)}=-{1\over 2}\epsilon^{abc}\nabla_b\xi_c.
\end{equation}
We had not spotted this simpler form, which is equal to our expression
\eqref{VOmega}.

\subsection{Rotating perfect fluid matter}

The stress-energy tensor for a perfect fluid is 
\begin{equation}
T_{ab}=(\rho+P)u_{a}u_{b}+Pg_{ab},
\end{equation}
where $u^a$ is tangential to the fluid worldlines, with $u^au_a=-1$,
and $P$ and $\rho$ are the pressure and total energy density measured
in the fluid frame. In the following, we assume the 1-parameter family
of ultrarelativistic fluid equations of state $P=\kappa\rho$, where
$0<\kappa<1$. In particular, $\kappa=1/2$ represents a 2-dimensional
gas of massless (or ultrarelativistic) particles in thermal
equilibrium, where the stress-energy tensor is trace-free. The sound
speed is $c_s=\sqrt{\kappa}$. There is no conserved rest mass density.

Following the Valencia formulation \cite{FontLRR08, Alcubierre08}, we
parameterize the 3-velocity $u^a$ in terms of the 2-velocity $v^a$ with
respect to a time slicing $t$ as
\begin{eqnarray}
\label{abstract3velocity}
u^a&:=&\Gamma(n^a+v^a), \\
\Gamma&:=&-n_au^a, \\
v^an_a&:=&0,
\end{eqnarray}
where $n^a$ is the future-pointing unit normal on the time slices.
The normalization $u^au_a=-1$ relates the Lorentz factor $\Gamma$ to the
2-velocity as 
\begin{equation}
\label{abstractGamma}
\Gamma^{-2}=1-v_av^a.
\end{equation}

Following standard practice in fluid dynamics in curved spacetime, we
write the stress-energy conservation equation $\nabla_aT^{ab}=0$ as a
set of three balance laws
\begin{equation}
\nabla_a\left(V_{b(i)}T^{ab}\right)=T^{ab}\nabla_{(a}V_{b)(i)},
\label{precovariantbalancelaws}
\end{equation}
or
\begin{equation}
\label{covariantbalancelaws}
\nabla_aj_{(i)}^a=s_{(i)},
\end{equation}
specified by a choice of three vector fields $V^a_{(i)}$. We have
already defined the vector fields $V_{(Z)}^a$ and $V_{(\Omega)}^a$,
which give rise to conservation laws (balance laws with zero source
term), and so are natural choices.

For the radial momentum (force) balance law we choose
\begin{equation}
V_{(Y)}^a:=\nabla^a(\ln R).
\end{equation}
This is the only choice where the resulting balance law is
``well-balanced'' for a fluid of constant density at rest in Minkowski
spacetime, in the sense that the flux term is constant and the source
term vanishes. By contrast, a balance-law based on any other choice of
$V_{(Y)}^a$ requires an explicit cancellation of the flux and source
terms, which may lead to large and unnecessary numerical error. An
equivalent choice for the radial momentum balance law was made in
\cite{Montero14} for spherical polar coordinates in $3+1$ dimensions
(without restriction to spherical symmetry).

\section{Description in polar-radial coordinates}
\label{section:section3}

\subsection{Metric and Einstein equations}

We now introduce a specific coordinate system, namely the
generalized polar-radial coordinates $(t,r,\theta)$, in terms
of which the axisymmetric metric takes the form
\begin{eqnarray}
\label{trmetric}
ds^2&=&-\alpha^2(t,r)\,dt^2+a^2(t,r)R'^2(r)\,dr^2
\nonumber \\ &&
+R^2(r)[d\theta+\beta(t,r)\,dt]^2.
\end{eqnarray}
Note that our choice $g_{rr}=a^2R'^2$ makes $a$ invariant under a
redefinition $r\to \tilde r(r)$ of the radial coordinate. 
The volume forms are given by
\begin{equation}
\epsilon_{tr\theta}=\alpha a R'R, \qquad
\bar\epsilon_{tr}=\alpha a R', 
\end{equation}
where we have made a choice of overall sign.

We assume that the spacetime has a regular central world line
$R=0$, and there we impose the gauge conditions, $\alpha(t,0)=1$,
$\beta(t,0)=0$, and the regularity condition $a(t,0)=1$. The gauge is
fully specified only after also specifying the strictly increasing
function $R(r)$, but we shall always assume that $R(r)$ is an odd
analytic function with $R(0)=0$, $R'(0)=1$.
The Killing vector is
\begin{equation}
\xi^a=\left({\partial\over\partial\theta}\right)^a
\end{equation}
and $R$ is its length, as above. We define the auxiliary quantity
\begin{equation}
\label{gammadef}
\gamma:=\beta_{,r},
\end{equation}
anticipating that $\beta$ will not appear undifferentiated in the Einstein
or fluid equations, but only in the form of $\gamma$ and its
derivatives, since the form (\ref{trmetric}) of the
metric is invariant under the change of angular
variable $\theta \to \theta + f(t)$.

Polar-radial coordinates have been used successfully in studying
critical collapse in spherical symmetry in $3+1$ spacetime dimensions,
starting with \cite{Choptuik93}. Their main advantage is that they
allow a fully constrained formulation of the Einstein equations, where
at $t=0$ and each subsequent timestep we solve differential equations
for $a$, $\alpha$ and $\beta$ that contain only $r$-derivatives. Their
main disadvantage is that they are apparent-horizon avoiding: in
spacetime regions where an apparent horizon is about to form, the
lapse $\alpha$ collapses near the center compared to its value far out
so that the time slicing stops advancing near the center and never
reaches the apparent horizon. This means that we cannot look very far
into black holes.

In our coordinates, $J$ and $M$ are given by
\begin{eqnarray}
J(t,r)&=&{R^3\gamma\over R'a\alpha}, \label{Jdefcoords} \\
M(t,r)&=&{R^2\over\ell^2}+{J^2\over 4 R^2}-{1\over a^2}. \label{Mdefcoords}
\end{eqnarray}

In an axistationary vacuum ansatz, $M$ and $J$ are constant in space and
time with value equal to the BTZ parameters of the same name. The BTZ
2-parameter family of metrics \cite{Banados92} takes the form
\begin{eqnarray}
\alpha^2&=&-M+{R^2\over\ell^2}+{J^2\over 4R^2}, \\
a^2 &=&{1\over \alpha^2}, \\
\beta&=&-{J\over 2R^2},
\end{eqnarray}
in all BTZ solutions.
The anti-de Sitter solution (from now, adS3) in particular is given by
$M=-1$ and $J=0$. Note that $\alpha a=1$ in the BTZ solutions.

In contrast to higher dimensions, stationarity actually follows from
vacuum axisymmetry locally, intuitively because there are no gravitational waves
in $2+1$ dimensions. The situation in $2+1$ axisymmetry is therefore
rather more similar to spherical symmetry in higher dimensions, where
the vacuum solutions are static and characterized by only a mass parameter.

Each BTZ solution is in fact locally, although not globally,
isometric to the adS3 solution \cite{BHTZ93}. However, this additional
symmetry will not be apparent in what follows.

The matter and Einstein equations are simplest in the standard
polar-radial coordinates defined by $R(r)=r$. However, in these
coordinates the coordinate speed of ingoing and outgoing radial light
rays is $dr/dt=\pm \lambda_c$, where
$\lambda_c:=\alpha/(aR')$. This increases rapidly with radius in
the BTZ solution, even in adS3. A necessary stability condition for
any numerical method for evolving ultrarelativistic fluid matter is
the Courant-Friedrichs-Levy (from now on, CFL) condition that the
numerical grid be wider than the light cones, that is $\Delta r/\Delta
t\ge \lambda_c$, everywhere in spacetime. As we require $R_{\rm
max}\gg \ell$ in situations of physical interest, this makes for a
wastefully small $\Delta t$.

This problem is easily fixed if we introduce
compactified polar-radial coordinates \cite{Bizon11}
\begin{equation}
\label{compactifiedR}
R(r)=\ell\tan(r/\ell),
\end{equation}
where the radial coordinate now has the range $0\le r<\ell\pi/2$.
In a vacuum region $\rho = 0$, where the metric is BTZ, the light
speed then takes the form
\begin{equation}
\lambda_c = 1-
	\left(1+M-\frac{J^2}{4 R^2}\right) \cos^2 r/\ell.
\end{equation}
In particular, the light speed is always bounded above and below.
In the adS solution, we have $\lambda_c=1$, and the CFL
condition is uniform. Similarly, the coordinate light speed will
remain bounded in asymptotically adS3 solutions. In our numerical
simulations we use the compactified coordinates (\ref{compactifiedR}),
with different values of the cosmological scale $\ell$, but for clarity
we will write $R$ and $R'$ rather than the explicit expressions.

Of the six algebraically independent components of the Einstein
equations in generalized polar-radial coordinates, five can be solved
for $\gamma_{,r}$, $\gamma_{,t}$, $a_{,r}$, $a_{,t}$ and
$\alpha_{,r}$. The undifferentiated shift $\beta$ does not appear in
the Einstein equations or in our formulation of the matter
equations. The sixth Einstein equation is a combination of first
derivatives of the other ones, and so is redundant modulo
stress-energy conservation.

To write the first four Einstein equations (\ref{DaJ},\ref{DaM}) in
coordinates, we define the current components
\begin{eqnarray}
Z&:=&\sqrt{-g}\,j_{(Z)}^t, \\
f_{(Z)}&:=&\sqrt{-g}\,j_{(Z)}^r, \\
\Omega&:=&\sqrt{-g}\,j_{(\Omega)}^t, \\
f_{(\Omega)}&:=&\sqrt{-g}\,j_{(\Omega)}^r,
\end{eqnarray}
and obtain
\begin{eqnarray}
\label{dJdr}
J_{,r}&=&16\pi Z, \\
\label{dJdt}
J_{,t}&=&-16\pi f_{(Z)}, \\
\label{dMdr}
M_{,r}&=&16\pi \Omega, \\
\label{dMdt}
M_{,t}&=&-16\pi f_{(\Omega)}.
\end{eqnarray}
The resulting conservation laws (\ref{DajaZ},\ref{DajaOmega}) take the form
\begin{eqnarray}
\label{ZFZcons}
Z_{,t}+f_{(Z),r}&=&0, \\
\label{OmegaFOmegacons}
\Omega_{,t}+f_{(\Omega),r}&=&0. 
\end{eqnarray}

A useful choice for the fifth independent Einstein equation, which
must contain $\alpha_{,r}$ in order to be independent of
(\ref{dJdr}-\ref{dMdt}), is
\begin{equation}
\label{dlnalphaaoRplphadr}
(\ln \alpha a)_{,r}=8\pi a^2RR'(1+v^2)\sigma,
\end{equation}
as the right-hand side vanishes in vacuum. The matter quantities $v$
and $\sigma$ in the right-hand side of this equation will be
defined below.

The Einstein equations (\ref{dJdr}-\ref{dMdt}) and
\eqref{dlnalphaaoRplphadr} are all linear combinations of components of
the Einstein equations, and so contain the fluid density, pressure and
velocity undifferentiated. We have not used the contracted Bianchi identities
(stress energy conservation), two of which are separately given as
(\ref{ZFZcons}-\ref{OmegaFOmegacons}).

\subsection{Balance laws}

Rather than working directly with the coordinate components $v^r$ and
$v^\theta$ of the 2-velocity, we use its frame components
in the radial and tangential directions,
\begin{equation}
v:=aR' v^r, \qquad w:=R v^\theta.
\end{equation}
We define the 2-velocity to be analytic if in the Cartesian coordinates
$x:=R\cos\theta$ and $y:=R\sin\theta$, its Cartesian components $v^x$
and $v^y$ are analytic functions of $x$ and $y$. This is the case in
axisymmetry if and only if $v$ and $w$ are analytic odd functions of
$R$, and hence of $r$ (as we choose $R(r)$ to be analytic and odd).

In terms of $v$ and $w$, and with $n_\mu=(-\alpha,0,0)$, the
3-velocity (\ref{abstract3velocity}) of the fluid is
\begin{equation}
u^\mu=\{u^t,u^r,u^\theta\}=\Gamma\left\{\frac{1}{\alpha},\frac{v}{aR'},
\frac{w}{R}-\frac{\beta}{\alpha } \right\},
\end{equation}
or equivalently
\begin{equation}
u_\mu=\Gamma \left\{-\alpha +R w \beta,aR'v,R w\right\},
\end{equation}
where the Lorentz factor (\ref{abstractGamma}) is
\begin{equation}
\Gamma^{-2}=1-g_{ij}v^iv^j=1-(v^2+w^2). \label{Gammadef}
\end{equation}

In coordinates, the balance laws take the form
\begin{equation}
\left(\sqrt{-g}\,V_{\mu(i)}T^{t\mu}\right)_{,t}
+\left(\sqrt{-g}\,V_{\mu(i)}T^{r\mu}\right)_{,r}=\sqrt{-g}s_{(i)}.
\end{equation}
We abbreviate this as
\begin{equation}
{\bf q}_{,t}+{\bf f}_{,r}={\bf S}.
\end{equation}
Note that the factor $\sqrt{-g}=\alpha a R'R$ is included in our
definitions of the conserved quantities ${\bf q}$, fluxes ${\bf f}$
and sources ${\bf S}$, and hence they depend on the choice of
coordinates, while the currents $j^a_{(i)}$ and sources $s_{(i)}$
in (\ref{covariantbalancelaws}) are defined
covariantly by (\ref{precovariantbalancelaws}).

The coordinate components of the three vector fields are
\begin{eqnarray}
V_{(Z)}^\mu&=&\{0,0,1\}, \\
V_{(X)\mu}&=&\{{\alpha\over a},0,0\}, \\
V_{(Y)\mu}&=&\{0,{R'\over R},0\}.
\end{eqnarray}
Note these do not all have the index in the same position --- we have
chosen the simplest form. 
The corresponding three balance laws have the conserved quantities
\begin{equation}
{\bf q}:=\{\Omega,Y,Z\}
\end{equation}
given by
\begin{eqnarray}
\label{Xdef}
X&=& R'R \tau, \\
Y&=& R'v \sigma, \\
Z&=& a R^2R' w \sigma, \\
\Omega&=&X+{JZ\over 2R^2},
\label{OmegafromX}
\end{eqnarray}
with the corresponding fluxes ${\bf f}$ given by
\begin{eqnarray}
\label{fX}
f_{(X)}&=& {\alpha\over a} R v \sigma, \\
f_{(Y)}&=& {\alpha\over a}(P+v^2\sigma), \\
f_{(Z)}&=& \alpha R^2 v w \sigma, \\
\label{fOmega}
f_{(\Omega)}&=&f_{(X)}+{Jf_{(Z)}\over 2R^2},
\end{eqnarray}
and the corresponding sources ${\bf S}$ by 
\begin{eqnarray}
\label{SX}
S_{(X)}&=& {1\over a}\Bigl[
-Rv\sigma\alpha (\ln a\alpha)_{,r}
\nonumber \\ &&
+R^2vw\sigma\gamma -RR'(1+v^2)\sigma a_{,t} 
\Bigr] \\
\label{SXEins}
&=& {1\over a}R^2vw\sigma\gamma={R'\over R ^3}Jf_{(Z)}, \\
\label{SY}
S_{(Y)}&=& {1\over a}\Bigl[
(w^2-v^2)\sigma\alpha {R'\over R} -\tau\alpha_{,r} \nonumber \\ &&
-(P+v^2\sigma)\alpha(\ln a)_{,r} \nonumber \\ &&
+Rw\sigma\gamma -2v\sigma R' a_{,t} \Bigr], \\
\label{SZ}
S_{(Z)}&=&0, \\
S_{(\Omega)}&=&0,
\label{SOmega}
\end{eqnarray}
where we have defined the shorthands
\begin{eqnarray}
\label{sigmadef}
\sigma&:=&\Gamma ^2 (1+\kappa) \rho, \\
P&:=&\kappa\rho, \\
\tau&:=&\sigma-P.
\label{taudef}
\end{eqnarray}
Note that in flat spacetime $S_{(X)}$ vanishes and only the
first term in $S_{(Y)}$ is present.

The specific metric derivatives appearing in $S_{(X)}$ and
$S_{(Y)}$ are given by the Einstein equations as
\begin{eqnarray}
\label{dlnalphadr}
(\ln \alpha)_{,r}&=& a^2RR'\left(
8\pi(P+v^2\sigma)-{J^2\over 4R^4}+{1\over \ell^2}\right),\nonumber \\ \\
\label{dlnadr}
(\ln a)_{,r}&=&a^2RR'\left(
8\pi\tau+{J^2\over 4R^4}-{1\over \ell^2}\right), \\
\label{dadt}
a_{,t}&=&- 8\pi\alpha a^2Rv\sigma.
\end{eqnarray}

In (\ref{SXEins}), we have used (\ref{dlnalphaaoRplphadr}) [which
itself follows from (\ref{dlnalphadr}) and (\ref{dlnadr})] and
(\ref{dadt}) to simplify $S_{(X)}$ to something that is proportional
to $J$ and so vanishes in spherical symmetry. In (\ref{SOmega}), we
have used the Einstein equations (\ref{dJdr},\ref{dJdt}) as well as
the conservation laws for $X$ and $Z$. By contrast, there is no
particular simplification when the Einstein equations are used to
express the metric derivatives in $S_{(Y)}$ in terms of the
stress-energy.

\subsection{Characteristic velocities}

The coordinate characteristic velocities $\lambda=dr/dt$
of the matter are the
eigenvalues of the $3\times3$ matrix $\partial {\bf f}/\partial {\bf q}$.
It is useful to
write the latter as $(\partial{\bf q}/\partial {\bf u})^{-1}
(\partial{\bf f}/\partial {\bf u})$, where as our primitive variables
we choose
\begin{equation}
{\bf u}:=\{\rho,v,w\}. 
\end{equation}
We find the coordinate characteristic velocities 
\begin{align}
\lambda_{0, \pm}&={\alpha\over aR'} \left\{
v,\ {v(1-\kappa)\Gamma^2\over(1-\kappa)\Gamma^2+\kappa} \right. \nonumber\\
& \left. \pm{\sqrt{\kappa(1-\kappa)(1-v^2)\Gamma^2+\kappa^2} 
	\over(1-\kappa)\Gamma^2+\kappa} \right\}
\end{align}
These represent the radial fluid velocity and the velocity of outgoing
and ingoing sound waves (in axisymmetry in $2+1$ dimensions, there are
only radial sound waves). In the (unphysical) limit $\kappa=1$, the
two sound velocities $\lambda_\pm$ reduce to $\pm\lambda_c$,
the coordinate speed of radial light rays. However, the fluid motion
will in general become relativistic even for $c_s=\sqrt{\kappa}\ll 1$,
and so $v$ will approach $\pm 1$ arbitrarily closely, which then means
that one of $\lambda_+$ approaches $\lambda_c$ or $\lambda_-$
approaches $-\lambda_c$.

\section{Numerical method}
\label{section:section4}

\subsection{Fluid evolution}

We use standard finite-volume methods for the time evolution of the
fluid variables. We initially discretize only in $r$. Time will be
discretized at the end, an approach sometimes called the method of
lines. We use standard notation where $r_i$ denotes cell centers and
$r_{i+1/2}$ denotes cell faces. In principle, each cell is
allowed to have a different width, but we always have
\begin{equation}
r_i:={1\over 2}(r_{i-1/2}+r_{i+1/2}). 
\end{equation}
We define the shorthand
\begin{equation}
\Delta_i(r):=r_{i+{1\over2}}-r_{i-{1\over2}},
\end{equation}
and similarly for other grid functions.

The numerical values of the conserved variables represent cell
averages (denoted by an overbar), that is
\begin{equation}
\bar{\bf q}_i(t):={1\over \Delta_i(r)}
	\int_{r_{i-\frac{1}{2}}}^{r_{i+\frac{1}{2}}}{\bf q}(t,r)\,dr
\end{equation}
in terms of notional continuum functions ${\bf q}(t,r)$. They are
updated by notional fluxes through cell faces plus notional cell
averages of the source terms, that is
\begin{equation}
{d\bar{\bf q}_i\over dt}={1\over \Delta_i(r)}
	\left({\bf f}_{i-{\frac{1}{2}}}-{\bf f}_{i+{\frac{1}{2}}}\right)+
		\bar {\bf s}_i. \label{dqdt}
\end{equation}
This update is conservative by construction when the source terms
vanish, simply because the fluxes from adjacent cells cancel in the
time derivative of $\int{\bf q}\,dr$. 

In the numerical code, where array indices must be integers, we label
cell $i$ by array index $i$ (obviously) and cell-face $r_{i+1/2}$ by
$i$, so each cell face is labeled by the cell to its left. The
physical cells are labeled $i=1,\dots N$ and their boundaries
$i=0,\dots N$, with $r_{1/2}:=0$ labeled as cell face $0$.

To find the numerical fluxes, we first reconstruct the fluid variables
in each cell in order to find left and right values at the cell
faces. In the reconstruction we use a slope limiter such as centered,
minmod or van Leer's MC limiter \cite{vanLeer73}. This takes as its
input the cell average of the conserved quantity, as well as some
slope information.

For these and other standard reconstruction methods to work well, the
functions ${\bf w}$ we reconstruct should be ``generic'' in the sense
that if we only have the cell average our best guess for the
reconstructed function should be constant over the cell (with value
equal to the cell average). However, none of our conserved quantities
and not all of our primitive variables are generic in this sense, as
they are expected to vary as some power of $R$ near the symmetry
boundary $R=0$. In particular, $v$ and $w$ are odd functions of
 $R$ (or $r$). By contrast, the functions we reconstruct are chosen
to be even functions of $R$ (or of $r$) that generically do not vanish
at $R=0$ (or $r=0$), namely 
\begin{eqnarray}
\label{wdef}
{\bf w}&:=&(\omega,\eta,\zeta) :=\left\{{\Omega\over
	R'R},{Y\over R'R},{Z\over R'R^3}\right\} \\
&=&\left\{\tau+{J\over2}{a w\sigma\over R},{v\sigma\over
	R},{a w\sigma\over R}\right\}.
\label{wfromqtilde}
\end{eqnarray}
We now approximate $\omega$, $\eta$ and $\zeta$ as constant in each
cell to find their notional cell center values ${\bf w}_i$
from the cell averages of the ${\bf q}$. For such functions, ${\bf
 w}(r)\simeq {\bf w}_i\simeq \bar {\bf w}_i$ is the best
approximation to make inside the $i$th cell whereas for a function
that behaves like a power of $R$ at the center it would not be. For
example, from (\ref{wdef}) we have
\begin{equation}
\omega\,d\left({R^2\over 2}\right)=\Omega\,dr.
\end{equation}
Approximating $\omega(r)=\omega_i$ and integrating over the $i$th
cell, and similarly for $\eta$ and $\zeta$, we obtain
\begin{eqnarray}
\label{omegafromOmega}
\omega_i &=& {2\Delta_i(r) \over \Delta_i (R^2)} \bar\Omega_i, \\
\eta_i &=& {2 \Delta_i(r) \over \Delta_i (R^2)} \bar Y_i, \\
\zeta_i &=& {4 \Delta_i(r) \over \Delta_i (R^4)} \bar Z_i.
\label{zetafromZeta}
\end{eqnarray}
We use these cell center values ${\bf w}_i$ together with
notional slopes to reconstruct ${\bf
w}(r)$ to the cell faces and, independently, the ${\bf w}_i$
(only) to compute the source terms at the cell centers.

To find the numerical fluxes ${\bf f}_{i+1/2}$, we approximate the
reconstruction as constant on each side of a cell face and then
solve the resulting Riemann problem. Note that to find the flux
through the cell face we do not need the
complete solution of the Riemann problem but only the value ${\bf
q}(r_{i+1/2})$ at the cell face. As the solution of the Riemann
problem is self-similar, 
\begin{equation}
{\bf q}(t,r)=\tilde{\bf q}\left({r-r_{i+1/2}\over t-t_n}\right),
\end{equation}
${\bf q}(t,r_{i+1/2})$ is time-independent, and so therefore is
${\bf f}_{i+1/2}:={\bf f}[\tilde{\bf q}(0)]$.

In practice, we do not solve the Riemann problem exactly but use an
approximate Riemann solver. We use the very simplest one, the HLL
approximate Riemann solver (\cite{Einfeldt88}). This approximates the
solution as a two-shock solution with shock speeds given \textit{a priori}
as $\pm\lambda_{\rm HLL}$. Conservation then forces the middle state to
be the average of the left and right state, and the resulting HLL
flux is given by
\begin{equation}
{\bf f}_{i-\frac{1}{2}} = \frac{{\bf f}({\bf q}^R_{i-1}) + 
	{\bf f}({\bf q}^L_{i}) + \lambda_{\rm HLL} \left({\bf q}^R_{i-1} -
	{\bf q}^L_{i}\right)}{2}, \label{HLLflux}
\end{equation}
where ${\bf q}^R_{i-1}$ and ${\bf q}^L_{i}$ are the right and left
reconstructions in the $(i-1)$th and $i$th
cells. $\lambda_{\rm HLL}$ is an estimate of the absolute value of
the largest coordinate characteristic speed. We use the coordinate
speed $\lambda_c$ of radial light rays, which is a (sharp) upper limit
for the matter characteristic speeds.

We impose regularity boundary conditions at the center by using
ghost points and the fact that all our grid functions are either even
or odd in $r$. We fill the outer ghost cells by extrapolating the
${\bf u, \bar{q}}$ or ${\bf w}$ as constant functions (copy boundary conditions).

We found some obstacles in extending the numerical outer boundary
to infinity. The HLL flux limiter is not positivity preserving, which
can lead to unphysical values for the density during the evolution. This
is offset by imposing a numerical floor (typically $\sim 10^{-14}$).
When extending the numerical grid to infinity, the outer boundary is
typically a region of near vacuum, where the density is then set to this
floor value. During the RK steps, the numerical flux continuously attempts
to reduce the density below the floor value. The density is then replenished
back to the floor value, thus continually adding mass to the system.
It is possible to circumvent this problem by not imposing a floor
on the density. In parallel, one can modify the numerical flux to be
positivity preserving by ``interpolating'' between the HLL flux with
some other positivity-preserving flux (such as Lax-Friedrichs)
\cite{Patrick21}. Doing so however generates shocks near the boundary that
quickly grow and travel inwards. We have not attempted to further
investigate this issue.

\subsection{Recovery of primitive variables}
\label{section:cons2prim}

To recover the primitive variables ${\bf u}$ from the conserved
variables ${\bf q}$ at one point, we first convert the ${\bf q}$ to
the ${\bf w}$. We then compute
\begin{equation}
\label{taufromomega}
\tau=\omega-{J\zeta\over 2}.
\end{equation}
Inverting (\ref{sigmadef}-\ref{taudef},\ref{wdef}), we compute
\begin{eqnarray}
\label{rhofromtau}
\rho&=&\frac{\tau}{[\Gamma^2 (1+\kappa)-\kappa]},
\\
\label{vfrometa}
v&=&\frac{R\eta}
{\Gamma^2 (1+\kappa )\rho},
\\
\label{wfromzeta}
w&=&\frac{R\zeta}
{a\Gamma^2 (1+\kappa)\rho}.
\end{eqnarray}
The Lorentz factor $\Gamma$ can be written in terms of $\bf w$, by plugging
(\ref{vfrometa},\ref{wfromzeta}) into \eqref{Gammadef} and solving for $\Gamma$.
We find
\begin{equation}
\label{GammafromU}
\Gamma^2=\frac{1-2\kappa(1+\kappa)U+\sqrt{1-4\kappa U}}
{2\left(1-(1+\kappa)^2U\right)},
\end{equation}
where we defined
\begin{eqnarray}
\label{Udef}
U&:=&{R^2(\eta^2+{\zeta^2\over a^2})\over (1+\kappa)^2\tau^2} \\
&=&\frac{\Gamma^2 \left(\Gamma^2-1\right)}{\left[
	\Gamma^2 (1+\kappa )-\kappa\right]^2}.
\end{eqnarray}

Note that the ${\bf w}$ must obey the constraint
\begin{equation}
R^2\left(\eta^2+{\zeta^2\over a^2}\right)<\tau^2
\label{geneconstraint}
\end{equation}
for the fluid velocity to be physical (timelike). Numerical error may lead to
this condition being violated, in which case (\ref{GammafromU}) fails.

\subsection{Einstein equations, fluxes and sources}

We need to already have the metric coefficients $J$ and $a$ (as well
as the given functions $R$ and $R'$) to recover the primitive
variables from the conserved variables, and in addition we need
$\alpha$ to compute the fluxes and sources. Moreover, variables can be
represented numerically as cell-center values, cell-face values, or
cell averages. Taking all this into account, in our fully constrained
evolution scheme we interleave the solution of the Einstein equations
at constant $t$ with the recovery of the primitive variables in the
following order, see also Table~\ref{table:numericalscheme} for a summary.

0) We start with the cell averages $\bar{\bf q}_i:=(\bar\Omega_i,\bar
Y_i,\bar Z_i)$ at some moment of time.

1) We find the cell-center values ${\bf w}_i:=(\omega_i,\eta_i,\zeta_i)$ using
(\ref{omegafromOmega}-\ref{zetafromZeta}).

2) We now come to the first of two blocks of metric calculations. We
find $J$ and $M$ at the cell faces by integrating out from $J=0$ and
$M=-1$ at the cell face $r=0$, using
\begin{eqnarray}
\Delta_i(J)&=&16\pi \bar Z_i\Delta_i r, \\
\Delta_i(M)&=&16\pi \bar \Omega_i\Delta_i r.
\end{eqnarray}
These integrals are exact as $\bar \Omega_i$ and $\bar Z_i$ represent
cell averages. As $\Omega$ and $Z$ are conserved exactly by our numerical
scheme this discretization also gives us exact conservation of $J$ and
$M$. From $J$, $M$ and $R$ at the cell faces we find $a$ 
at the cell faces using (\ref{Mdefcoords}).

$a$ is a generic even function, so using the average of the values at
the two cell faces is a reasonable approximation to its value at the
cell center,
\begin{equation}
a_i={1\over 2}(a_{i-1/2}+a_{i+1/2}). 
\end{equation}

At the same time, we determine $\tau$ at the cell centers. This is more
subtle, as it involves $Z$ and $J$, which scale as $Z\sim R^3$ and
hence $J\sim R^4$ near the center and so are not generic even
functions. We first approximate $Z$ in cell $i$ by assuming that
$\zeta$, which is a generic even function, is constant in the cell (at
the cell-center value $\zeta_i$, which we found from the cell average
$\bar Z_i$). This gives the approximation
\begin{equation}
\label{Zapprox}
 Z_i\simeq{4\Delta_i(r)\bar Z_i\over\Delta_i(R^4)}R^3_i R_i'.
\end{equation}
We also have the exact relation
\begin{equation}
J(r_i)=J_{i-1/2}+16\pi\int_{r_{i-1/2}}^{r_i}Z(\tilde r)\,d\tilde r
\end{equation}
and an equivalent expression integrating from $r_{i+1/2}$. Inserting
the approximation (\ref{Zapprox}), carrying out the integration, and
averaging the two resulting expressions for $J(r_i)$, we find the
approximation
\begin{equation}
\label{Japprox}
J_i\simeq {\Sigma_i(J)\over 2}+8\pi\Delta_i(r)\bar
	Z_i{2R^4_i-\Sigma_i(R^4)
		\over\Delta_i(R^4)},
\end{equation}
where 
\begin{equation}
\Sigma_i(J):=J_{i-1/2}+J_{i+1/2}
\end{equation} 
and similarly for other grid functions.

We evaluate the approximation (\ref{Japprox}) at
the cell centers to obtain $J_i$, and hence $\tau_i$. 

3) We now have $\tau_i$, $\eta_i$ and $\zeta_i$ and the metric
coefficient $a_i$ at the cell centers, and recover the primitive
variables ${\bf u}_i:=(\rho_i,v_i,w_i)$ at the cell centers as
described in Sec.~\ref{section:cons2prim}.

4) We now come to a second block of metric calculations. We integrate
the remaining Einstein equation (\ref{dlnalphaaoRplphadr}) in the
approximation
\begin{equation}
\label{lnalphaadiscrete}
\Delta_i\left(\ln(\alpha a)\right)\simeq 4\pi
a_i^2(1+v_i^2)\sigma_i\Delta_i(R^2)
\end{equation}
to obtain $a\alpha$ and hence $\alpha$ at the cell faces,
starting from the gauge condition $\alpha(t,0) = 1$.

We interpolate $\alpha$ to the cell centers, as we did for $a$. From
$J$, $a$ and $\alpha$ we compute $\gamma$ at the cell faces and cell
centers using (\ref{Jdefcoords}). As a diagnostic only, we find
$\beta$ at the cell faces by integration using the trapezoid rule, and
then interpolate $\beta$ to the cell centers. We start the
integration of $\beta$ from the gauge condition $\beta(t,0) = 0$.

5) We evaluate (\ref{dlnalphadr}-\ref{dadt}), and hence (\ref{SY}) at
the cell centers to find the source term $S_{(Y)i}$ at the cell
centers. As $S_{(Y)}\sim RR'f$, where $f$ is a generic even function
near the center, we integrate the approximation $f_i=\bar f_i$ over
the $i$-cell to find
\begin{equation}
\label{SYbar}
\bar S_{(Y)i}={S_{(Y)i}\over R_iR_i'}{\Delta_i(R^2)\over2\Delta_i(r)}.
\end{equation}

6) We use a standard slope-limited method to reconstruct the ${\bf
	w}$ to the cell faces, denoting the value immediately to the left of
the cell face at $r_{i-1/2}$ by ${\bf w}_{i-1}^R$ and the value
immediately to the right by ${\bf w}_{i}^L$. We already have values of
$J$ and $a$ at the cell faces (continuous across the cell face). We
find $\tau$ at both sides of each cell face using
(\ref{taufromomega}), $U$ from (\ref{Udef}), then $\Gamma$ and finally
the ${\bf u}$. Finally, we use an approximate Riemann solver to find
the numerical fluxes ${\bf f}$ through the cell faces from the ${\bf
	u}$ on each side.

7) We then have $d\bar{\bf q}_i/dt$ from (\ref{dqdt}).

\begin{table}
	\renewcommand\arraystretch{1.5} \centering
	\caption{\label{table:numericalscheme} Overview of how $d{\bf q}/dt$
		is calculated. Steps 2 and 4 are not required if the
                metric is fixed. ``+ floor'' means that we
                  impose a floor on small quantities at this point.}
	\begin{tabular}{r|c|c}
		0) & $\bar{\bf q}_i:=(\bar\Omega_i,\bar Y_i,\bar Z_i)$ + floor & 
		(\ref{Xdef}-\ref{OmegafromX})\\
		\hline
		1) & ${\bf w}_i:=(\omega_i,\eta_i,\zeta_i)$ + floor & (\ref{wdef})\\
		\hline
		2) & $J_{i+1/2}$, $M_{i+1/2}$, $a_{i+1/2}$ & (\ref{dJdr},\ref{dMdr},\ref{Mdefcoords}) \\
		& $a_i$ & average \\
		& $J_i$, $\tau_i$ & (\ref{Japprox},\ref{taufromomega}) \\
		\hline
		3) & ${\bf u}_i:=(\rho_i,v_i,w_i)$ + floor & 
		(\ref{Udef},\ref{GammafromU},\ref{rhofromtau}-\ref{wfromzeta})\\
		\hline
		4) & $\alpha_{i+1/2}$ & (\ref{lnalphaadiscrete})\\
		& $\alpha_i$ & average \\
		& $\gamma_i$, $\gamma_{i+1/2}$ & (\ref{Jdefcoords}) \\
		& $\beta_{i+1/2}$ & (\ref{gammadef}) \\
		& $\beta_i$ & average \\
		\hline
		5) & $\bar S_{(Y)i}$ via $S_{(Y)i}$ & (\ref{SY},\ref{SYbar}) \\
		\hline 
		6) & ${\bf f}_{i - 1/2}$ via ${\bf w}_i^L$, ${\bf w}_{i-1}^R$, ${\bf
			u}_i^L$
		${\bf u}_{i-1}^R$ & (\ref{fX}-\ref{fOmega},\ref{HLLflux}) \\
		\hline
		7) & $d\bar{\bf q}_i/dt$ & (\ref{dqdt}) \\
	\end{tabular}
\end{table}
\subsection{Imposition of a floor on small quantities}

Recall that the generic variables need to satisfy the constraint
\eqref{geneconstraint} everywhere at all times. Failure for this
condition to be satisfied results in an unphysical value of
\eqref{Udef} and thus of $\Gamma^2$. A primary concern is to ensure
that this inequality is satisfied in near-vacuum regions, since in
those regions all three of the variables $\tau, \eta, \zeta$ are small.
We choose to impose a floor on the generic variables at each physical cell,

\begin{equation}
\tau_i - R_i \sqrt{\eta_i^2 + {\zeta_i^2 \over a_i^2}} \ge \delta_\text{f}.
\label{numericalfloor}
\end{equation}
If the above condition is not satisfied at any cell $i$, we proceed as
follows: First, $\tau_i$ is set to be at least the floor value,
\begin{equation}
\tau_{i, \text{ new}} = \max \left(\delta_\text{f}, \tau_i \right).
\end{equation}

Then we split the density and momentum variables into an ingoing and
an outgoing combination (defined in the spirit of characteristic variables),
and impose a floor on each separately,
\begin{equation}
c_\pm := \max \left(\tau_i \pm R_i \sqrt{\eta_i^2 +
	{\zeta_i^2 \over a_i^2}}, \delta_\text{f}\right).
\end{equation}
Note that necessarily $c_- = \delta_\text{f}$.
The variables $\tau, \eta$ are then updated as,
\begin{align}
\tau_{i,\text{new}} &:= \frac{c_+ + c_-}{2},\\
\eta^2_{i,\text{new}} &:= \frac{(c_+ - c_-)^2}{4 R^2_i}-
	{\zeta_i^2 \over a_i^2}. \label{eta_new}
\end{align}
The sign of $\eta_{i,\text{new}}$ is chosen so that it has the same
sign as $\eta_i$. It is possible due to numerical errors that the rhs
of \eqref{eta_new} is negative. In this case, we set 
\begin{equation}
\eta_{i, \text{new}} = 0
\end{equation}
and solve \eqref{eta_new} for $\zeta_i \to \zeta_{i, \text{new}}$.
The updated value $\zeta_{i, \text{new}}$ can be written explicitly as
\begin{equation}
\zeta_{i,\text{new}} = 0
\end{equation}
if $c_+ = c_-$ and
\begin{equation}
|\zeta|_{i,\text{new}} = 
a \Bigl|{2 (\tau_i - \delta_\text{f}) \pm \sqrt{(\tau_i - 
	\delta_\text{f})^2 + 3 R_i^2 \eta_i^2} \over 3 R_i} \Bigr|,
\end{equation}
if $c_+ > c_-$. We select the root that minimizes $||\zeta|_i - 
|\zeta|_{i, \text{new}}|$ and again we choose the sign of
$\zeta_{i,\text{new}}$ to coincide with the sign of $\zeta_i$.

By construction, the updated values then satisfy \eqref{numericalfloor}.
The floor $\delta_\text{f}$ itself is computed as the maximum between a
relative and absolute floor,
\begin{equation}
\delta_\text{f} := \max \left(\delta_\text{abs}, \delta_\text{rel}
\left(\tau_i + R_i \sqrt{\eta_i^2 + {\zeta_i^2 \over a_i^2}}\right)\right).
\label{numericafloor}
\end{equation}
The addition of this second relative floor is due to the fact that it
is possible to encounter a situation for which $c_- < \delta_\text{f},
c_+ > \delta_\text{f}$ and also $c_+ \gg c_-$. In this case, within
numerical precision, the update of the generic variables do not register.
The second term in \eqref{numericafloor} ensures that the floor is never
``too small" compared to the data and that the update is therefore always
properly applied. Typical values we choose are $\delta_\text{abs} = 
\delta_\text{rel} = 10^{-12}$. The floor is applied to the generic variables
each time they are computed from the conserved variables. Furthermore,
within each Runge-Kutta step, the floor is imposed on the newly computed
conserved variables. This is done by first converting $\bf \bar q_i$ into
$\bf w_i$ using \eqref{omegafromOmega}-\eqref{zetafromZeta}, imposing the
floor on them as discussed above and then converting back to $\bf \bar q_i$
by inverting \eqref{omegafromOmega}-\eqref{zetafromZeta}. We note that each
time the floor is applied, the value of $\tau$ increases, resulting in the
associated conserved variables $\bar \Omega_i$ to also increase. Thus, due
to the floor, $\bar \Omega_i$ is not exactly conserved during the evolution.

\subsection{Overall time step and initial data}

Starting from the conserved quantities $\bar{\bf q}_i$ at one moment in
time we have now recovered the metric and primitive variables, and the
time derivative $d\bar{\bf q}_i/dt$. We implement (\ref{dqdt}) in a fourth
order Runge-Kutta scheme in $t$. Note that for high-resolution limiters such as
MC or minmod limiters, this scheme will also be total-variation-diminishing
\cite{Leveque02}. Each time we evaluate $d\bar{\bf q}_i/dt$ in the substeps
of that scheme we also recalculate the metric.

We impose symmetry boundary conditions at $r=0$, based on the fact
that all variables are either even or odd in $r$. As we start each
time step, and each Runge-Kutta timestep, assuming that only the $\bar
{\bf q}_i$ are known, we impose the symmetry boundary conditions on
them after each Runge-Kutta substep.

Any initial data in general relativity consist of a part that is
freely specified and a part that is obtained by solving the
constraints (and perhaps gauge conditions). As we have a fully
constrained scheme for solving the Einstein equations, it is natural to
prescribe the ``matter'' and use the Einstein equations to find the metric
coefficients, but the meaning of matter is necessarily ambivalent.
We specify the generic variables ${\bf w}_i$ at the cell centers as our free
initial data, from which we can immediately compute the averaged conserved
quantities ${\bf \bar{q}}_i$ from \eqref{omegafromOmega}-\eqref{zetafromZeta}.
From ${\bf \bar{q}}_i$, we can then follow the numerical scheme outlined in
Table~\ref{table:numericalscheme} to compute all the other quantities at the
initial time step in a consistent way. Note that specifying the ${\bf w}$, or
equivalently the ${\bf q}$, means that we know $M$ and $J$ \textit{a priori}.
This would not be the case if we specified the primitive variables ${\bf u}$.

\subsection{Formation of apparent horizon and computation of critical quantities}
\label{section:sub_super_check}

Since we are not using a horizon penetrating foliation, one cannot
observe the formation of an apparent horizon. We instead make use of
two simple criteria to determine if a given initial data will collapse
or disperse. For our intended application to critical collapse,
it is important that this decision can be reliably automated.

First, if during the evolution, the timestep
$\Delta t$ is smaller than some minimum timestep $\Delta
t_\text{min}$, then formation of apparent horizon is deemed to be
imminent and unavoidable and the corresponding initial data will be
judged as being supercritical. The rationale behind this is that the
time steps are computed so that the CFL condition is also satisfied,
\begin{equation}
\label{cCFL}
\Delta t = c_\text{CFL} \min_i \left(\Delta_i r\right) \min_{i, i-{1 \over 2}}
\left(\frac{a R'}{\alpha}\right),
\end{equation}
where the last minimum is computed from both the cell centers and
faces and $0 < c_\text{CFL} < 1$. It is well known that in spherical
symmetry, the formation of an apparent horizon is easily identified
with the vanishing of $(\nabla R)^2 = 1/a^2 = 0$ at some radius $R =
R_{AH}$. From the above and \eqref{dlnalphadr}-\eqref{dlnadr}, it
follows that the time step $\Delta t \to 0$ outside the horizon. A
typical value is $\Delta t_\text{min} = 10^{-11}$.

There are also two other
criteria that effectively act as fail-safes: if the maximum density
$\rho_\text{max}$ is larger than some threshold density at any point in time,
then this will also be deemed as supercritical data. A typical value is
$\rho_\text{threshold} = 10^{30}$. This criteria is usually never triggered
since the time step $dt$ becomes sufficiently small before this happens.

The second criterium is the value of $(\nabla R)^2$ itself. Since
on the onset of apparent horizon formation, $(\nabla R)^2 \to 0$, numerical
error can conspire to produce unphysical values of $(\nabla R)^2$, namely,
$(\nabla R)^2 \lesssim 0$. This will also be a sign that collapse is unavoidable.
If a given time evolution does not satisfy any of these criteria and the
evolution has run for a sufficiently long time, the initial data will be deemed
to be subcritical.

There is a subtlety in the notion of ``sufficiently long,'' in
that the negative cosmological constant effectively confines the
matter. For perfect fluid matter, this is due to an inward
cosmological acceleration. One may conjecture that, given enough
time, any initial data with total mass $M>0$ will form a black hole,
and this is well established numerically for scalar field matter
\cite{Bizon11}. As we impose an unphysical numerical
boundary condition at finite $R$, we are unable to investigate this,
and so our criteria are, in some sense, for prompt collapse.

To investigate scaling at the threshold of (prompt) collapse,
we need to record the maximum of the density
$\rho_\text{max}$ and the mass and spin of the apparent horizon $M_\text{AH},
J_\text{AH}$ respectively. The latter are computed using the formulas
\eqref{Jdefcoords} and \eqref{Mdefcoords} evaluated at the apparent horizon
$R_\text{AH}$. This is found from the minimum value of $(\nabla R)^2$,
$(\nabla R)^2_\text{min} := \min_{i,n} (\nabla R)^2_i(t_n)$ from which we then
consider the two neighboring points of $(\nabla R)^2_\text{min}$ and make a
polynomial interpolation. The variables needed in the computation of $M_\text{AH},
J_\text{AH}$ are then evaluated by linear interpolation from $R_\text{AH}$.

\section{Numerical tests}
\label{section:section5}

\subsection{Convergence testing}

In this section, we investigate the pointwise convergence as well
as convergence with respect to a norm of our numerical code for different
scenarios. Specifically, we examine six cases. First, we consider initial data
``far" from the black hole
threshold which disperses and collapses. For each of these two cases,
we will consider a ``slowly" and ``rapidly" rotating case. Finally, we
also consider initial data corresponding to rotating stars that
are presumed stable and unstable.

Let $f$ refer to any quantity of interest. In the following, we will
mostly be interested in the conserved variables $\bf \bar q$, as they
are used to evolve the data at the next timestep. It should still be
emphasized that the primitive and generic variables still indirectly
play a role in the evolution, notably during the floor imposition and
when computing the fluxes at the cell faces, see
Table~\ref{table:numericalscheme}. In our numerical code, we consider an
approximation to the exact function $f(t,r)$. This approximation
depends on the grid resolution $\Delta_i(r)$ and since we
always choose a uniform grid spacing in the simulations we may
simplify the notation by defining $h := \Delta_i(r)$. The
approximation of the exact solution $f(t,r)$ will then be denoted by
$F_h(t,r)$. The function $F_h(t,r)$ converges pointwise to the exact
solution $f(t,r)$ if at all points we have
\begin{equation}
F_h(t,r) = f(t,r) + C(t,r) h^k+ \mathcal{O}(h^{k+1}),
\end{equation}
where $C(t,r)$ is a smooth function which depends on the continuum
solution $f(t,r)$ and $k$ is the order of convergence. Typically, the
exact solution $f$ is unknown, but this problem can be circumvented by
considering instead the difference between two resolutions,
\begin{equation}
\delta F_h(t,r) := F_h(t,r) - F_{h \over 2}(t,r).
\end{equation}
It follows that our scheme converges to order $k$ if
\begin{equation}
\delta F_h(t,r) = 2^k \delta F_{h \over 2}(t,r) \left(1+ \mathcal{O}(h)\right).
\label{pointwise_convergence}
\end{equation}

Besides investigating pointwise convergence, we will also be
interested in the convergence in a norm. Consider the $\ell^2$ norm,
defined at any fixed time $t$ by
\begin{align}
||F||_2^2 (t;h,p) = {h \over 2} \sum_{i=1}^{N-p} 
	&\left(F_h(t, r_{i-1/2})^2 \right. \nonumber \\
&\left. + F_h(t, r_{i+1/2})^2\right).
\end{align}
Note that we use the cell faces instead of the cell centers, because
the former align exactly when we double the resolution.
If $F$ corresponds to fluid variables, such as $\bf u, {\bar q}$ or $\bf w$,
the cell faces values are computed from the cell centers by linear interpolation.

Recall that the center is located at $r_{1/2} = 0$, while the outer boundary
corresponds to $r_{N+1/2} =: r_\text{max}$. Note that in the definition of the
norm, we also allow the truncation of the last $p$ grid points for reasons that
will be explained shortly.

Applying this norm to \eqref{pointwise_convergence}, we then find that
\begin{equation}
\mathcal{N}_F(t;h,p) := \log_2 \left(||\delta F||_2(t;h,p) \over 
	||\delta F||_2(t;{{h \over 2},p})\right) = k + \mathcal{O}(h).
\end{equation}
By construction, one expects second-order convergence everywhere, except at
and near the outer boundary due to the copy boundary conditions. On the other
hand, the boundary conditions at the center are expected to not spoil the
second-order convergence since they preserve the even/oddness of the functions
they are applied to. 

In the following, we investigate the following points: First, the
correct implementation of the code, which should imply second-order
convergence at least at short times everywhere, except possibly near
the outer boundary. Second, we wish to investigate how the error that
originates from the boundary affects the inside of the numerical
grid. This is particularly important for the stationary
configurations, since the conserved quantities do \textit{not} vanish
at infinity and so one would a priori expect the numerical
outer boundary conditions
to play a crucial role. Pointwise convergence is useful as it can
highlight small numerical instabilities that would otherwise be
hidden when looking at the convergence in a norm. On the other hand,
convergence in a norm will be used to formalize the idea that the code
converges to order $k$ ``almost everywhere." Specifically, it is
possible that we find that some variables do not converge at all at
the boundary, but that these instabilities do not travel inside the
numerical grid, or if they do, they do it very slowly. In this case,
we then would expect $\mathcal{N}_F(h,0) \ll k$, while for some small
$p$, we would recover $\mathcal{N}_F(h,p) \simeq k$.

In what follows, we always consider the radiation fluid
equation of state $\kappa=1/2$. The numerical grid is equally spaced
in the compactified coordinate $r$, as defined in
\eqref{compactifiedR} and the Courant factor of (\ref{cCFL}) is
set to $c_\text{CFL}=0.5$. The cosmological constant is set to
$\Lambda = - \pi^2 / 4$, which sets the boundary of adS in
compactified coordinates to $r_\infty=1$.

\subsection{Dispersion and collapse}

For both dispersion and collapse, we consider the evolution of five
different grid resolutions, with $100\times 2^n$ points for $n$
from 1 to 5, so that for the lowest resolution, $h
\simeq 0.0035$. The numerical outer boundary is set at $r_\text{max} =
0.7$, corresponding to $R_\text{max} = \ell \tan(r_\text{max}/\ell) \simeq 1.25$,
and the copy boundary conditions will be imposed on the conserved variables. 

For slowly rotating dispersion and collapse, we will choose the
monotonized central-difference limiter (MC limiter) introduced by van
Leer \cite{vanLeer73}, while for the rapidly rotating cases, we
instead switch to a centered limiter, as the latter is empirically
found to be slightly more robust against numerical instabilities.
Independently, for rapidly rotating collapse the
convergence drops significantly at the onset of collapse. We found
that this can be partly offset by imposing no mass to enter the numerical
domain from the outer boundary by setting the HLL flux of $\Omega$ to be
zero if it is negative.

For dispersion, the simulation is stopped when most of the energy has
left the numerical domain, while for the case of collapse, we stop at
the onset of black hole formation, see Sec.~\ref{section:sub_super_check}.
We choose to initialize the generic fluid variables ${\bf w}$ as
double Gaussians in the area radius $R$,
\begin{align}
\omega(0,R) &= \frac{p_\omega}{2} \left(e^{-
			\left(\frac{R-R_\omega}{\sigma_\omega}\right)^2} + e^{-
			\left(\frac{R+R_\omega}{\sigma_\omega}\right)^2}\right), \label{initialdata_gaussian_omega} \\
\eta(0,R) &= \frac{p_\eta}{2} \left(e^{-
	\left(\frac{R-R_\eta}{\sigma_\eta}\right)^2} + e^{-
	\left(\frac{R+R_\eta}{\sigma_\eta}\right)^2}\right), \\
\zeta(0,R) &= \frac{p_\zeta}{2} \left(e^{-
	\left(\frac{R-R_\zeta}{\sigma_\zeta}\right)^2} + e^{-
	\left(\frac{R+R_\zeta}{\sigma_\zeta}\right)^2}\right),
\label{initialdata_gaussian_zeta}
\end{align}
where $p_\omega, p_\eta, p_\zeta$ are the magnitudes, $R_\omega,
R_\eta, R_\zeta$ the displacements from the center and $\sigma_\omega,
\sigma_\eta, \sigma_\zeta$ the widths of the Gaussians. For all four
cases, we set the widths to $\sigma_\omega = 0.2$,
$\sigma_\zeta = \sigma_\eta = 0.15$, and the
displacements to $R_\omega= R_\eta = R_\zeta = 0.4$. The slowly
rotating initial data have $p_\zeta = 0.01$, with
$p_\omega = 0.2$ for dispersion and $p_\omega = 0.5$ for
collapse. The rapidly rotating data have $p_\omega = 0.3$, $
p_\zeta = 0.5$, and $p_\omega = 0.5$ and $p_\zeta = 0.7$ for
dispersion and collapse respectively. In the ``slowly" and
``rapidly'' rotating data that collapse, the black hole mass and
spin parameter satisfy $J_\text{AH}/(M_\text{AH} \ell) \simeq
0.012$ and $0.9$ respectively. For all four test cases presented
above, the initial data satisfy the inequality \eqref{geneconstraint}
everywhere.

\begin{figure*}
	\includegraphics[width=2.2\columnwidth,height=1.1\columnwidth]
	{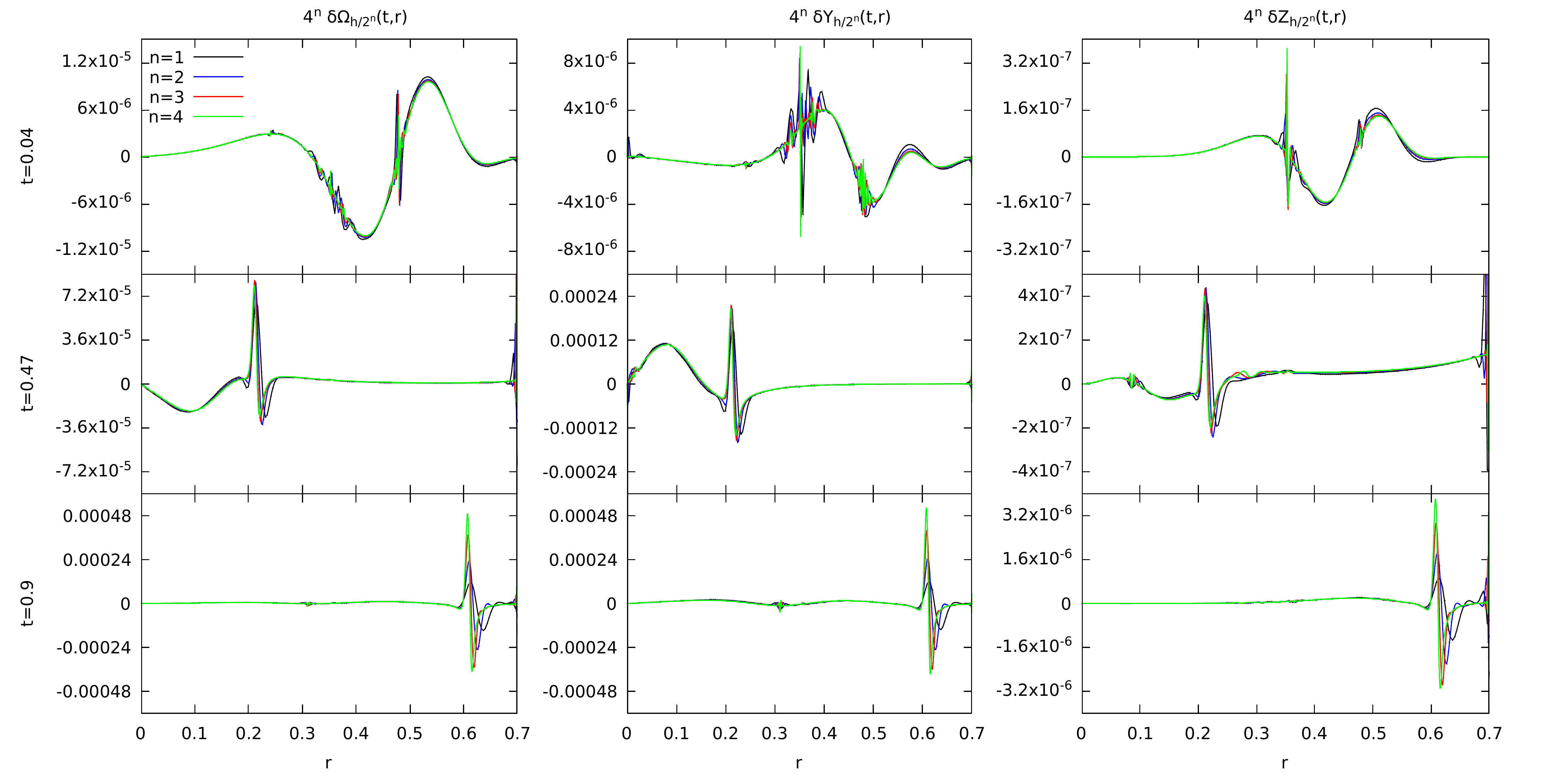}
	\caption{Dispersion with slow rotation: Plots of $4^n
		\delta {\bf \bar q}_{h \over 2^n}$ against $r$, at four
		resolutions $n=1,2,3,4$. From the left, the columns
		represent $\Omega$, $Y$ and $Z$, respectively, while the rows
		represent the times $t = 0.04$, $0.47$ and $0.90$, from the
		top. In each plot, the curves representing different
		resolutions approximately align, demonstrating pointwise
		second-order convergence. The unsmooth but convergent
		features of the error are artifacts of the MC limiter, and do not
		correspond to any visible unsmoothness of the solution itself. They
		are absent with the centered limiter.}
	\label{fig:dispersion_pointwise_convergence}
\end{figure*}

\begin{figure*}
	\includegraphics[width=2.2\columnwidth,height=1.1\columnwidth]
	{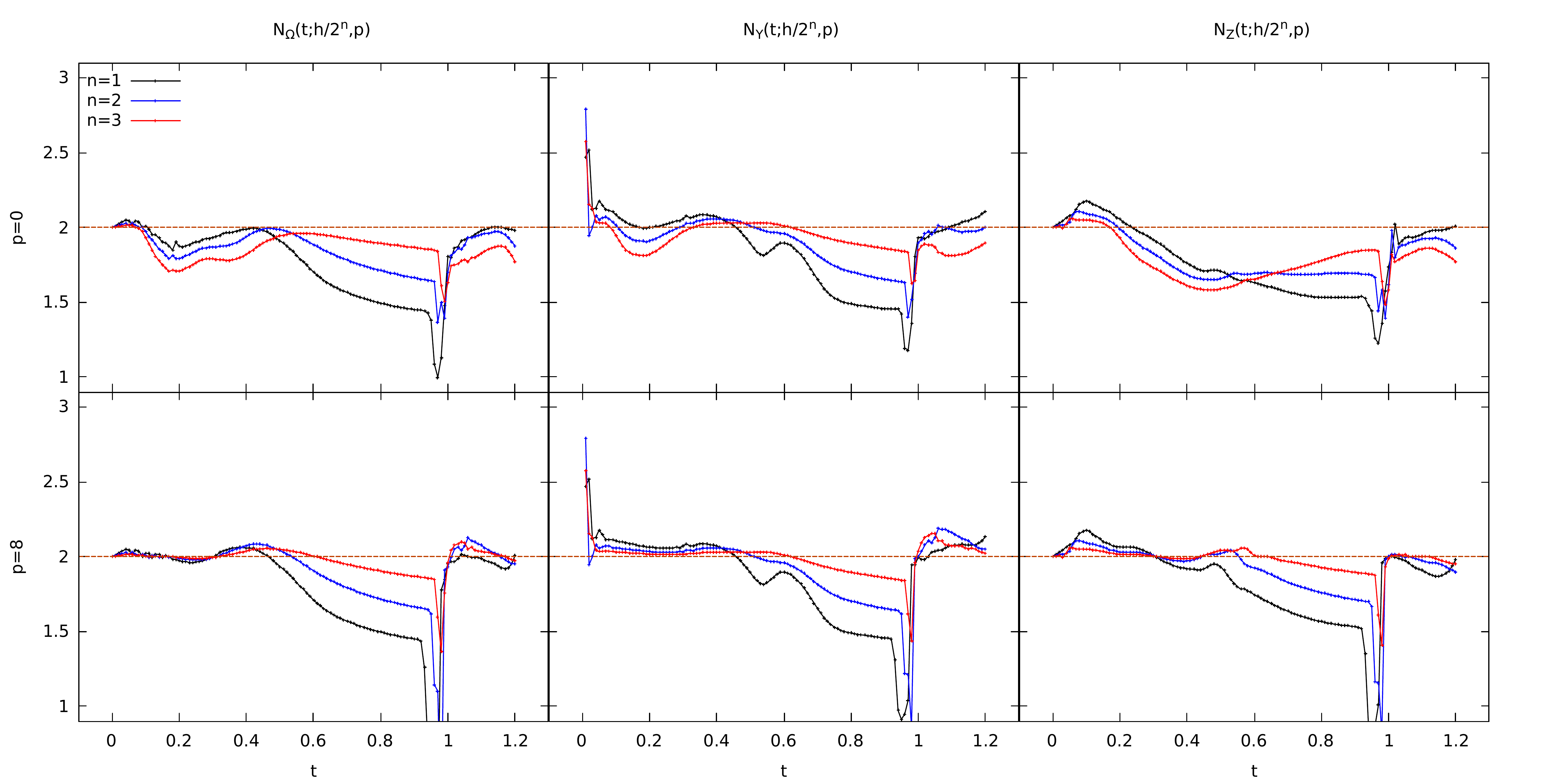}
	\caption{Dispersion with slow rotation: Plots of the
		convergence rates in the $\ell^2$-norm,
		$\mathcal{N}_{\bf \bar q} (t;{h \over 2^n},0)$ (upper row,
		all grid points used) and $\mathcal{N}_{\bf \bar q}(t;{h
			\over n},8)$ (bottom row, last 8 grid points omitted in
		the norm), for $n=1,2,3$. As in the previous
		figure the three columns represent $\Omega$, $Y$ and $Z$,
		respectively. The dashed horizontal line corresponds to
		second-order convergence, $\mathcal{N}=2$. When the full
		grid is taken into account in the computation of the norm,
		we typically observe less than second-order convergence. On
		the other hand, second-order convergence is recovered once
		the last 8 grid points are neglected in the computation of the norm.}
	\label{fig:dispersion_norm_convergence}
\end{figure*}

\begin{figure*}
	\includegraphics[width=2.2\columnwidth,height=1.1\columnwidth]
	{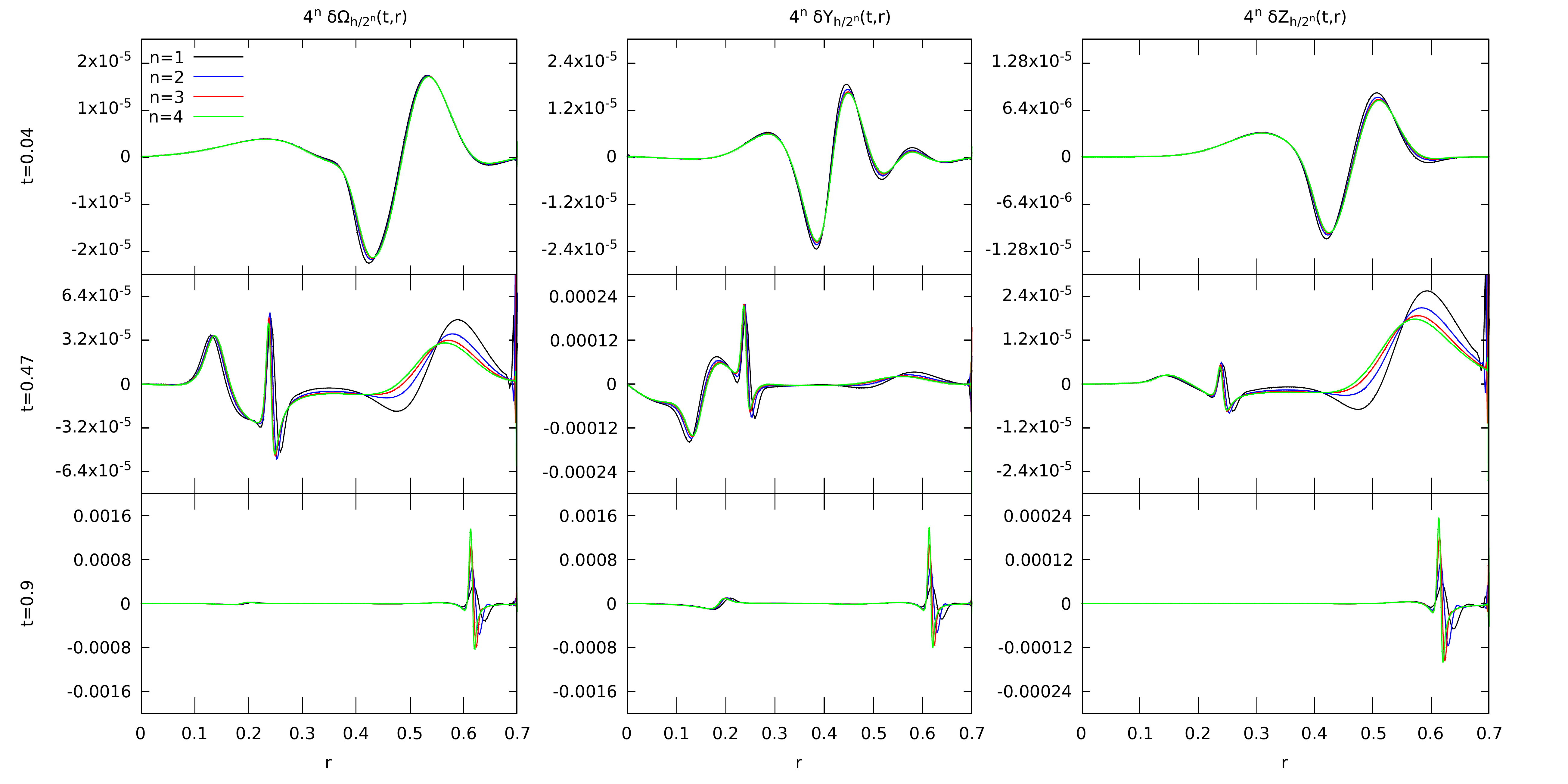}
	\caption{Dispersion with rapid rotation: Note that the instabilities in 
		Fig.~\ref{fig:dispersion_norm_convergence} at time $t = 0.04$ are not
		present here due to choosing a centered limiter instead of the MC limiter.
		Otherwise as in Fig.~\ref{fig:dispersion_pointwise_convergence}.}
	\label{fig:dispersion_high_rotation_pointwise_convergence}
\end{figure*}

\begin{figure*}
	\includegraphics[width=2.2\columnwidth,height=1.1\columnwidth]
	{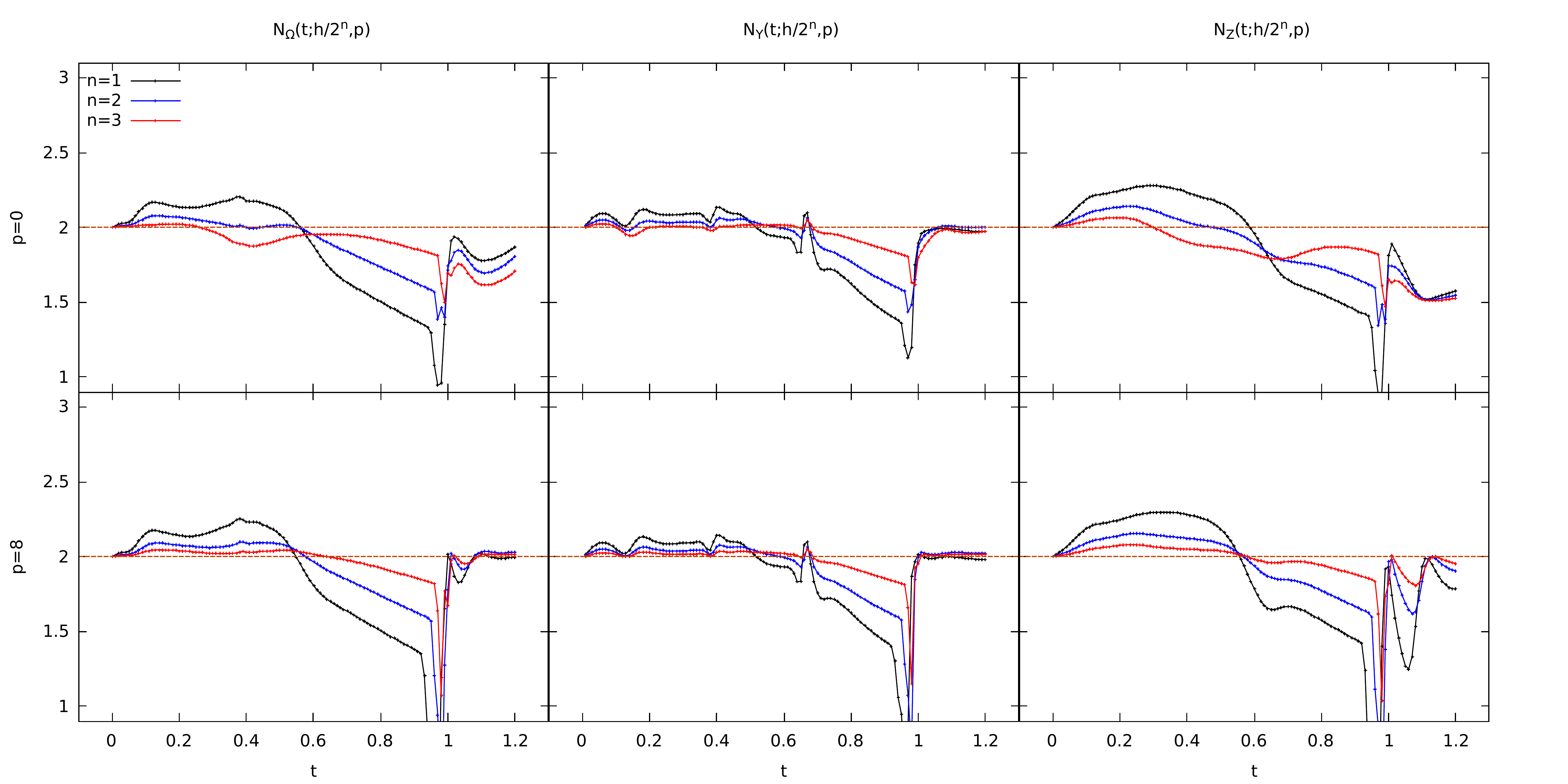}
	\caption{Dispersion with rapid rotation: Otherwise as
		in Fig.~\ref{fig:dispersion_norm_convergence}.}
	\label{fig:dispersion_high_rotation_norm_convergence}
\end{figure*}

In Fig.~\ref{fig:dispersion_pointwise_convergence}, we plot $4^n \delta
{\bf \bar q}_{h \over 2^n}$ (left, middle and right columns for $\Omega$,
$Y$ and $Z$, respectively) at four different resolutions $n=1,2,3,4$ for
initial data that disperses with small angular momentum. The profiles
are plotted at three different times (top, middle and bottom rows) $t
= 0.04, 0.36$ and $0.9$. These snapshots represent respectively, the
evolution of the error near the initial time, when the energy density
reaches a maximum (near the center), and when the matter finally
disperses and most of the density is about to leave the numerical
domain. During the evolution, the conserved variables remain smooth.

According to \eqref{pointwise_convergence}, the approximate alignment
of these plots shows that the code converges to second order. One can,
however, spot some instabilities at isolated points
inside the numerical grid. Their frequency increases with resolution,
but their amplitudes do not grow with time and in fact converge away
rather quickly with increased resolution. These instabilities are a
consequence of our choice of limiter as we observed that these
instabilities vanish with a centered limiter.

On the other hand, the convergence is mostly unaffected by the
choice of imposing copy boundary conditions on the
conserved variables instead of the primitive or generic
variables. Finally, as anticipated, we lose second-order convergence
at and near the outer boundary. The error propagates very slowly
inside the numerical domain and so does not spoil the second-order
convergence for most of the numerical grid for the period of time the
simulation is run.

To illustrate this, in Fig.~\ref{fig:dispersion_norm_convergence}
we plot $\mathcal{N}_{\bf \bar q}(t;{h \over 2^n},0)$ and
$\mathcal{N}_{\bf \bar q}(t;{h \over 2^n},8)$, for $n=1,2,3$. For the
former, untruncated case, we find that the order of convergence is
typically less than second order. On the other hand, we recover the
expected second-order accuracy once the last 8 grid points are ignored
in the calculation of the norm. The drop in convergence that can be
seen at around $t \simeq 1.0$ corresponds to the energy leaving the
numerical grid, see the last row of Fig.~\ref{fig:dispersion_pointwise_convergence}.

In
Figs.~\ref{fig:dispersion_high_rotation_pointwise_convergence} and 
\ref{fig:dispersion_high_rotation_norm_convergence},
we demonstrate second-order convergence pointwise and with respect to
the $\ell^2$ norm for the highly rotating dispersing initial data.
As for the slowly rotating case, the conserved variables remain
smooth during the evolution.

\begin{figure*}
	\includegraphics[width=2.2\columnwidth,height=1.1\columnwidth]
	{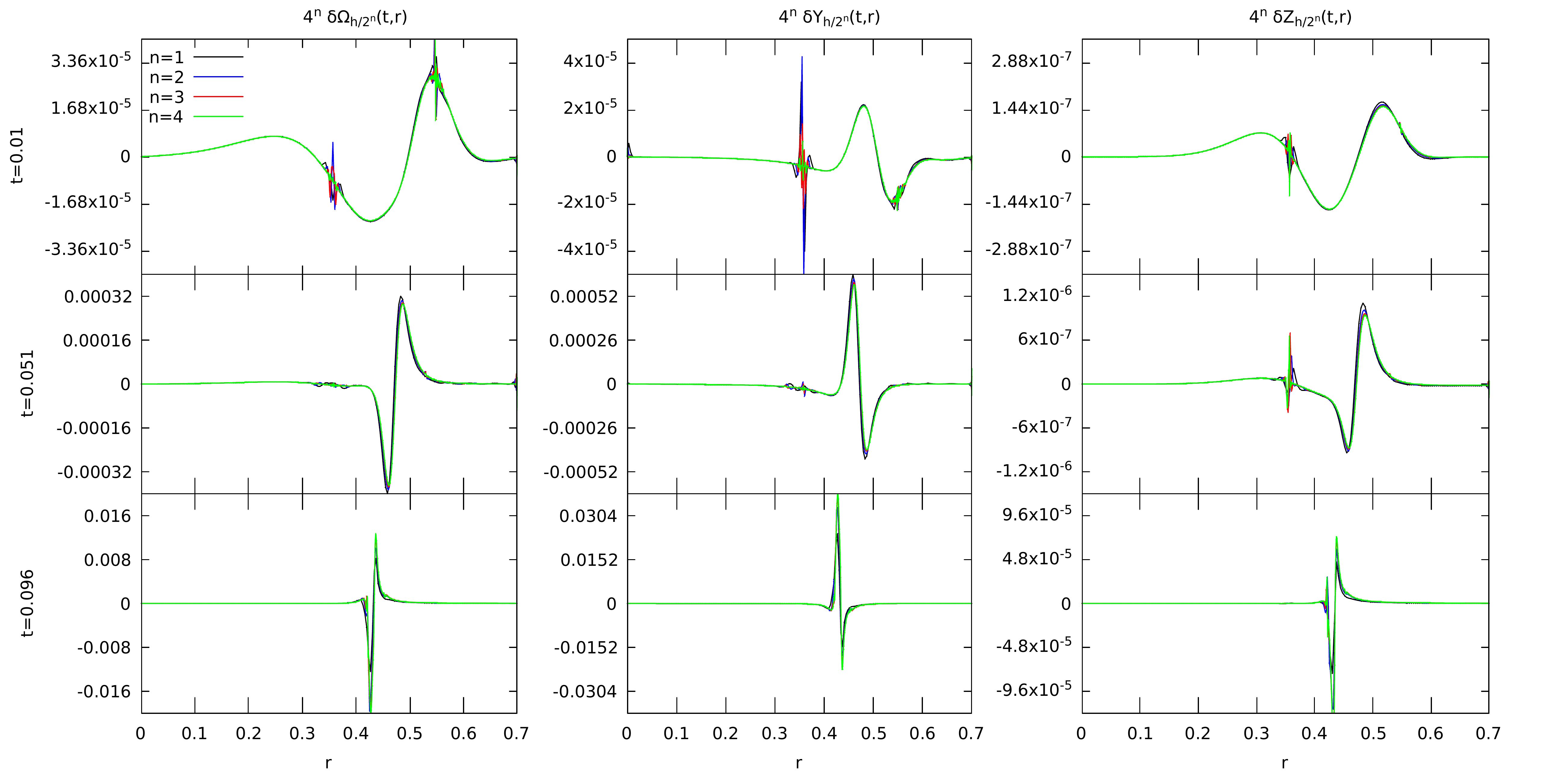}
	\caption{ Collapse with slow rotation: Times are now $t =
		0.010$, $0.051$ and $0.096$ (rows, from top to bottom), otherwise as in
		Fig.~\ref{fig:dispersion_pointwise_convergence}.}
	\label{fig:collapse_pointwise_convergence}
\end{figure*}

\begin{figure*}
	\includegraphics[width=2.2\columnwidth,height=1.1\columnwidth]
	{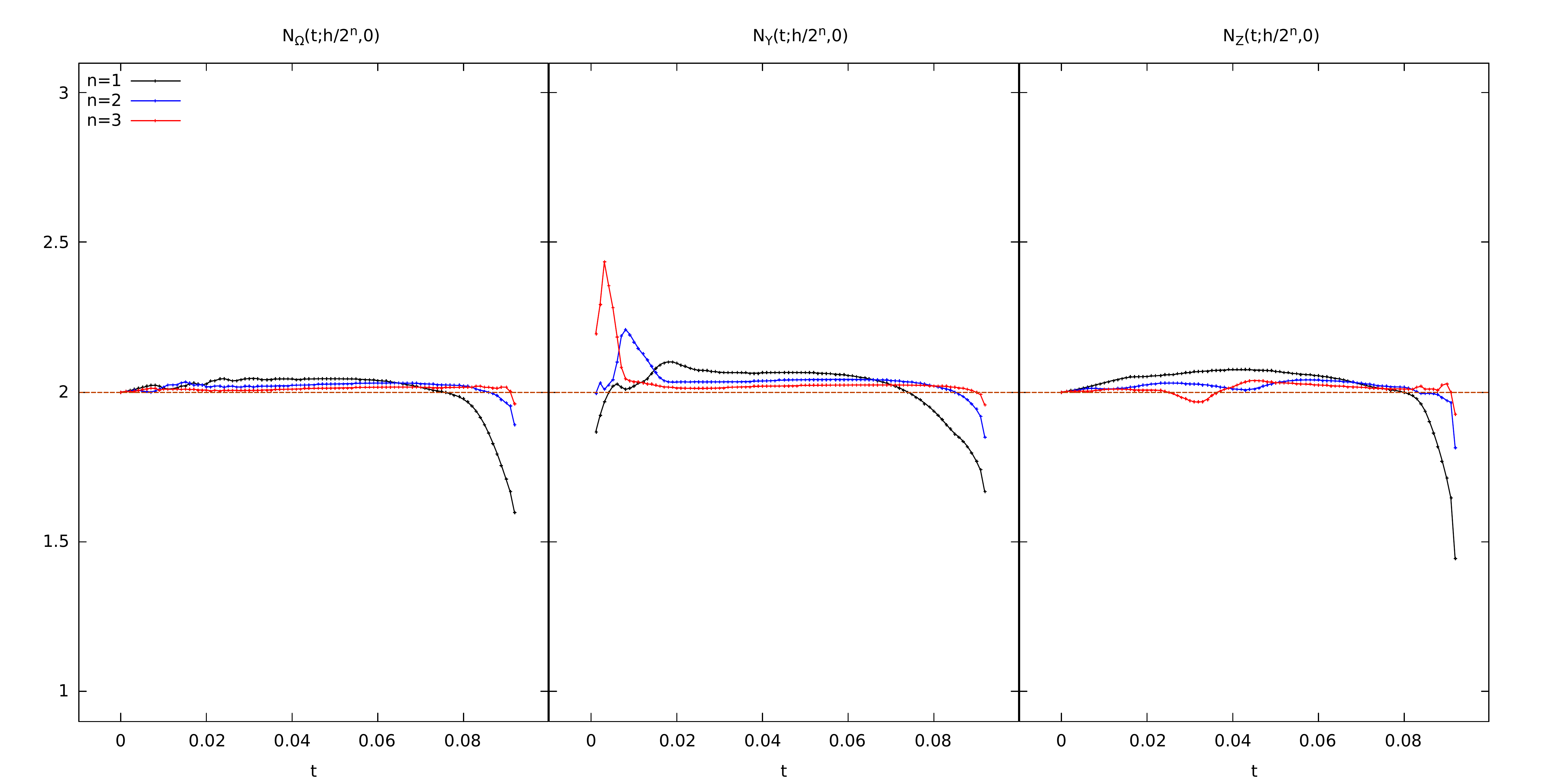}
	\caption{Collapse with slow rotation: Plots of $\mathcal{N}_{\bf \bar
			q}(t;{h \over 2^n},0)$, for $n=1,2,3$. As always, the columns correspond to
		$\Omega$, $Y$ and $Z$ from left to right. Due to the prompt collapse,
		second-order convergence is maintained throughout the evolution.}
	\label{fig:collapse_norm_convergence}
\end{figure*}

\begin{figure*}
	\includegraphics[width=2.2\columnwidth,height=1.1\columnwidth]
	{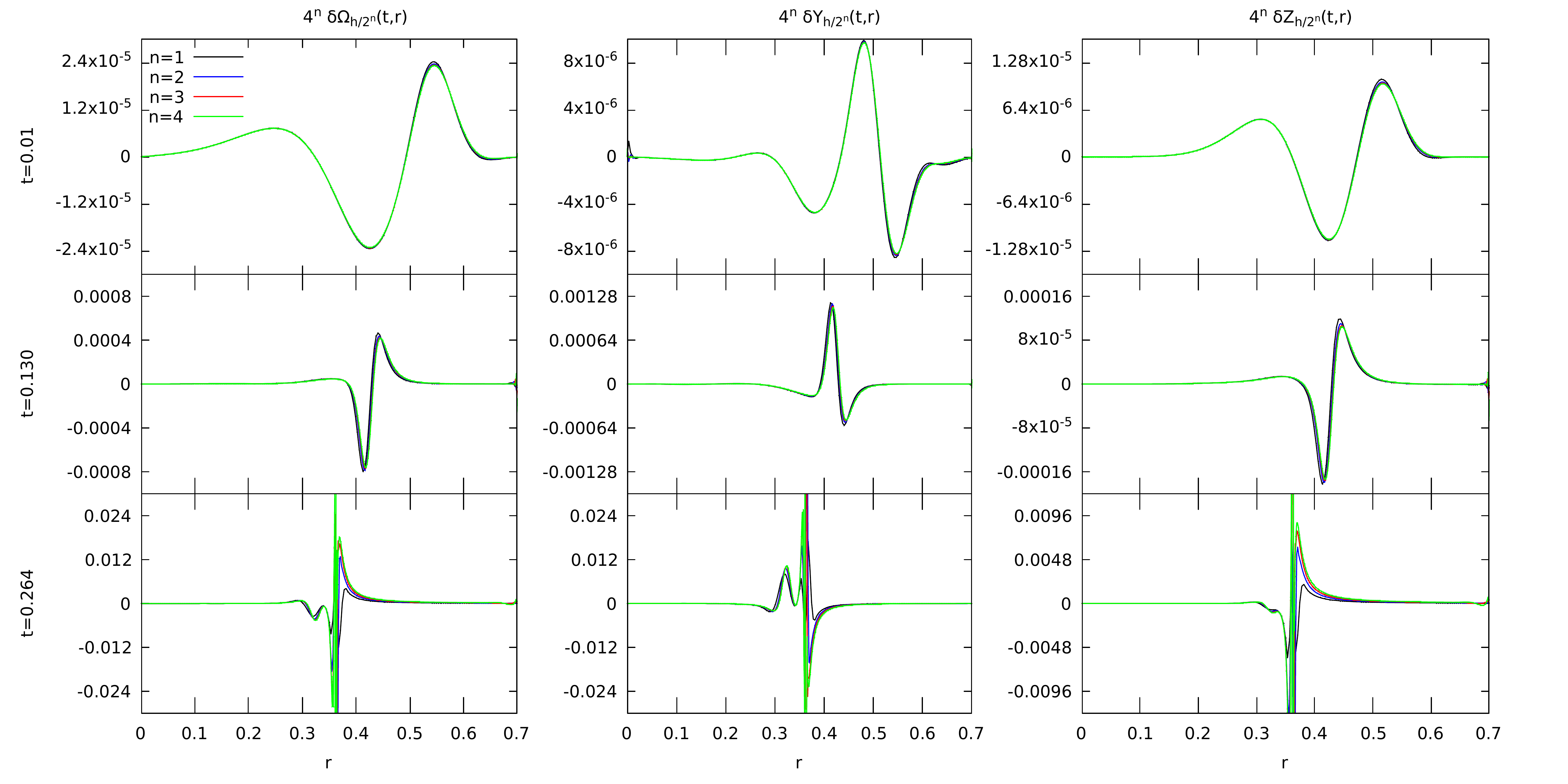}
	\caption{Collapse with rapid rotation: Times are now $t = 0.010$, $0.130$ and
		$0.264$, otherwise as in Fig.~\ref{fig:collapse_pointwise_convergence}.
		Note that we lose second-order convergence at the onset of collapse and
		near the region of black hole formation.}
	\label{fig:collapse_high_rotation_pointwise_convergence}
\end{figure*}

\begin{figure*}
	\includegraphics[width=2.2\columnwidth,height=1.1\columnwidth]
	{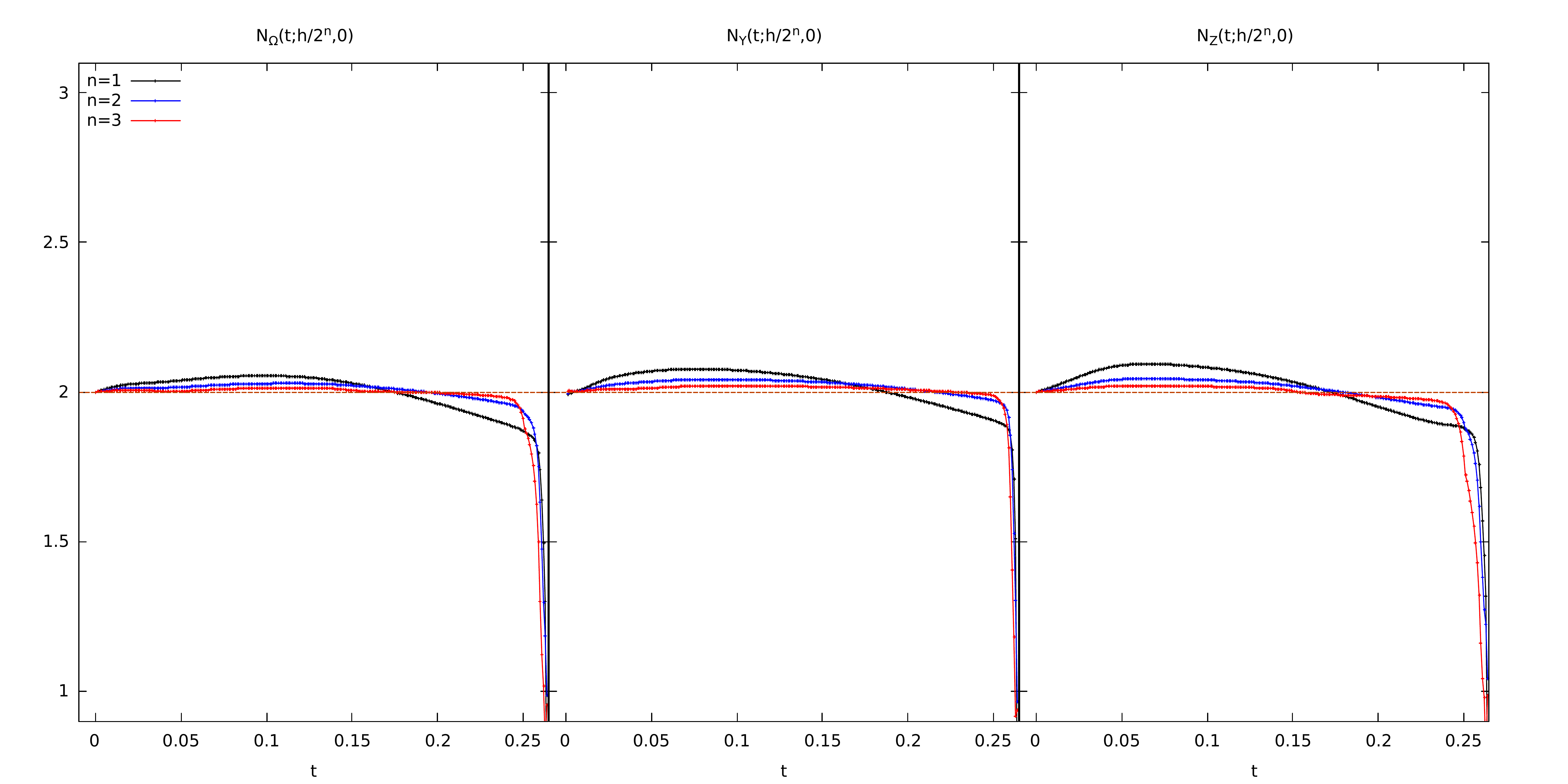}
	\caption{Collapse with rapid rotation: Otherwise as in
		Fig.~\ref{fig:collapse_norm_convergence}. Second-order
		convergence is lost near the onset of collapse, as
		seen also in the last row (time) of
		Fig.~\ref{fig:collapse_high_rotation_pointwise_convergence}.}
	\label{fig:collapse_high_rotation_norm_convergence}
\end{figure*}

Turning our attention now to the collapse case, in
Fig.~\ref{fig:collapse_pointwise_convergence} we show $4^n \delta
{\bf \bar q}_{h \over 2^n}$ at four different resolutions
$n=1,2,3,4$ at times $t=0.01, 0.051$ and $0.096$. We find the
same qualitative behavior as for the dispersion case, except
that the outer boundary behaves much better.

As a consequence, in Fig.~\ref{fig:collapse_norm_convergence}, we only
plot $\mathcal{N}_{\bf \bar q}(t;{h \over 2^n},0)$ as we have good
second-order convergence without the need to truncate the grid. As for
dispersion, the choice of limiter and which variables the outer
boundary conditions are applied to do not produce any qualitative
differences, except for the centered limiter which removes the 
instabilities already noted in the dispersion case, see
Fig.~\ref{fig:collapse_pointwise_convergence}.

Finally, in
Figs.~\ref{fig:collapse_high_rotation_pointwise_convergence} and
\ref{fig:collapse_high_rotation_norm_convergence},
we demonstrate second-order convergence for the case of rapidly
rotating collapsing data. As one would expect, the presence of angular
momentum delays the time of collapse. Near the onset of collapse the
convergence drops to first-order near the region where the horizon forms.

\subsection{Stable and unstable stars}

In \cite{Carsten20}, we analysed in detail the family of stationary
solutions parametrized by two dimensionless constants, $(\Omega_0,
\mu)$ or equivalently $(\tilde{J}, M)$, where we defined the
dimensionless spin
\begin{equation}
\tilde{J} := {J \over \ell}.
\end{equation}
In the parameter space $(\Omega_0, \mu)$, it was shown that the set of
parameters which result in a solution that is regular everywhere and
asymptotes to a BTZ solution with $\tilde{J} \leq M$ is doubly
covered for each admissible pair of values $(\tilde{J}, M)$. Both
regions are separated by a curve on which solutions have a
zero mode, i.e. a static linear perturbation that corresponds to an
infinitesimal change in $(\Omega_0, \mu)$ that leaves $(\tilde{J},M)$
invariant to linear order.

Such a double cover is familiar in $3+1$ dimensions, where the
less dense star is stable and the more dense star
unstable. Analogously, it was conjectured that the solution with the
smaller $\mu$ associated to a given $(\tilde{J},M)$ is
unstable, while the one with the larger $\mu$ is stable. We
use this opportunity to provide some numerical evidence for this
claim. Specifically, consider the pair of solutions with total mass
and angular momentum given by $\tilde{J} = 0.24, M = 0.38$,
corresponding to $(\Omega_0, \mu) \simeq (0.154, 0.242)$ and $(0.153,
0.392)$. These correspond to the black and orange dots in Fig. 1 in
\cite{Carsten20} and therefore to the unstable and stable
solutions associated to the above conserved quantities $\tilde{J},M$.

For both the stable and unstable configuration, 
we add a small Gaussian perturbation, with plus or minus sign.
The Gaussian perturbation is of the form
\eqref{initialdata_gaussian_omega}-\eqref{initialdata_gaussian_zeta},
with $|p_\omega|=0.001, p_\eta = p_\zeta=0$, $R_\omega=0.4$,
$\sigma_\omega=0.2$.
We set $r_\text{max}=0.9$ and consider again five different resolutions,
with the lowest resolution now $800$ grid points, or $h \simeq
0.00015$. We choose a larger value of $r_\text{max}$ because that the
stationary initial data under consideration do not have a surface at some finite
area radius. Consequently, one needs to choose a larger value of $r_\text{max}$
to fit ``most'' of the energy density inside the numerical grid.
We find that a MC or minmod limiter produces large
instabilities in the evolution and that these are mostly tamed with a
centered limiter. Furthermore, it is essential to use the primitive
variables for the copy boundary conditions. Using the
conserved variable instead causes the star to disperse almost
immediately due to a perturbation originating from the outer boundary,
while using the generic variables produces noticeably larger errors
during the evolution. We will therefore restrict to this choice in
what follows. Lastly, due to the nonvanishing of the conserved
variables at the boundary, it is necessary to impose, as for the highly
rotating collapse case, that the flux of $\Omega$ be non-negative at the
numerical outer boundary. For the stable stationary initial data, we also impose
the flux of $Z$ to be positive at the numerical outer boundary. (Note that
by construction, $Z$ is non-negative everywhere initially).

\begin{figure*}
	\includegraphics[width=2.2\columnwidth,height=1.1\columnwidth]
	{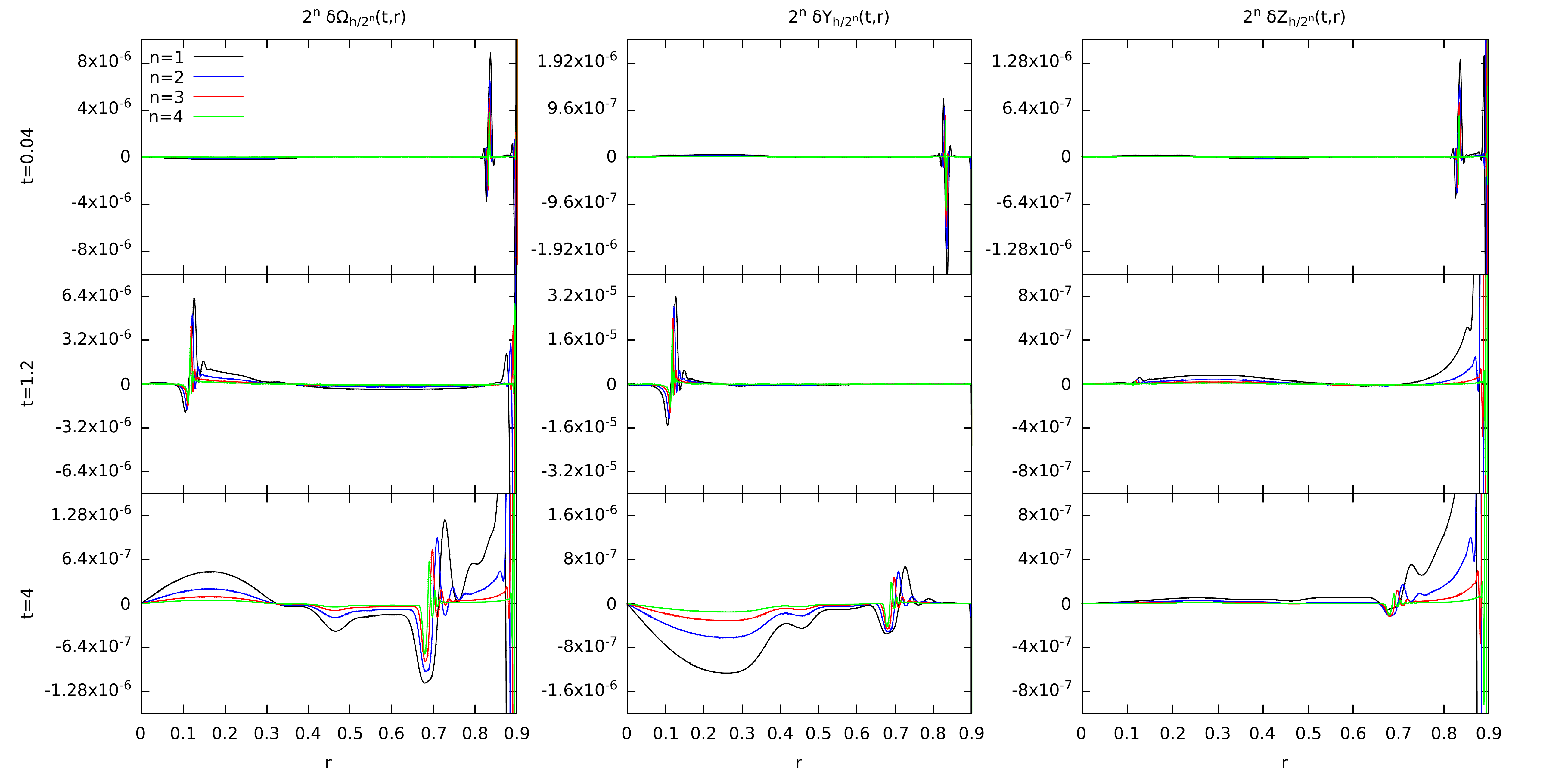}
	\caption{Stable stationary star: Times are now $t = 0.04$,
		$1.2$ and $4$ (rows, from top to bottom). We find $1 < \mathcal{N} < 2$
		inside the numerical grid. The numerical error is dominated by the
		outer boundary. This error does not converge, but travels
		inward very slowly and its width shrinks with increased resolution.}
	\label{fig:stable_pointwise_convergence}
\end{figure*}

\begin{figure*}
	\includegraphics[width=2.2\columnwidth,height=1.1\columnwidth]
	{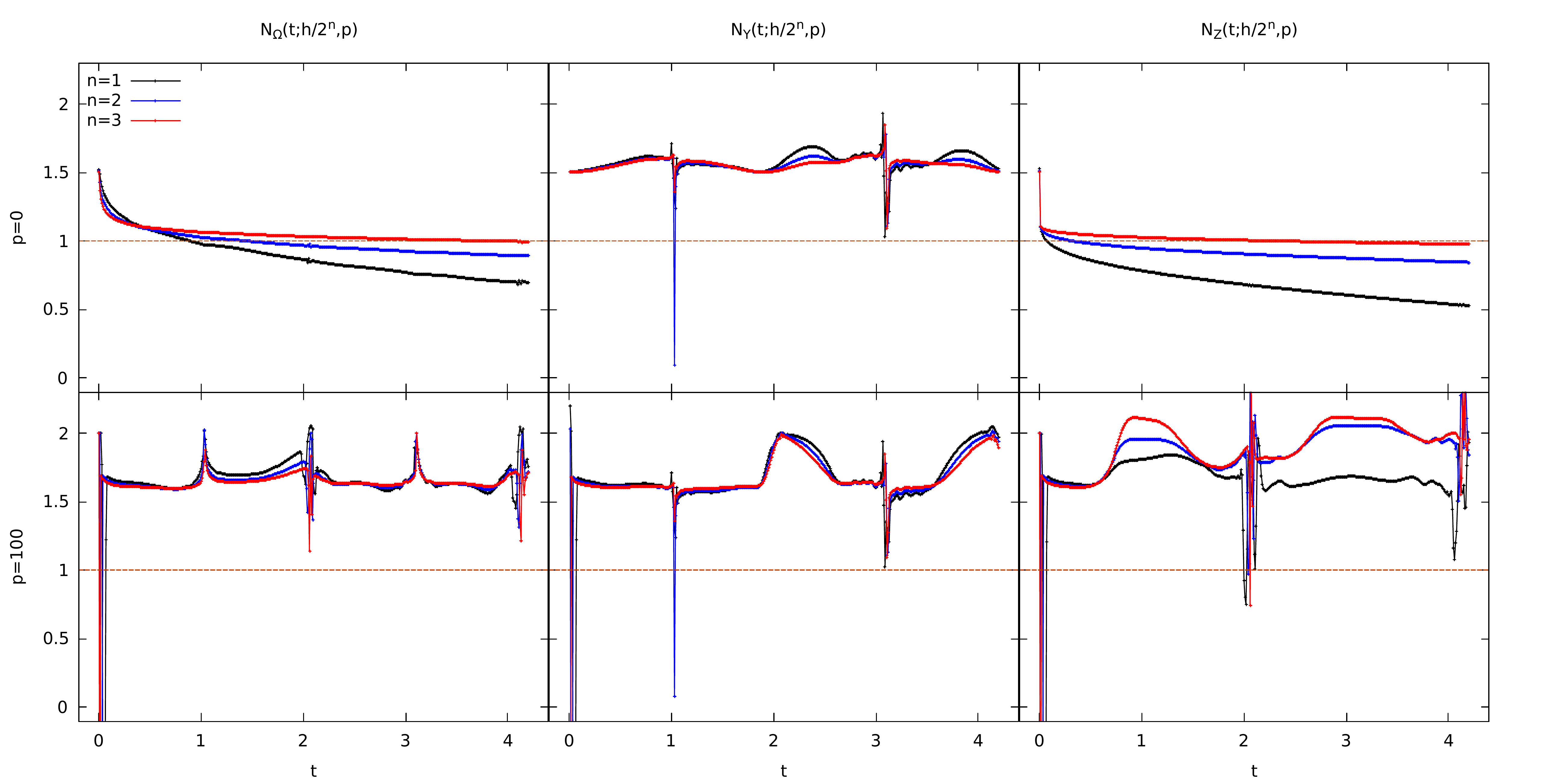}
	\caption{Stable stationary star: Plots of $\mathcal{N}_{\bf \bar q}
		(t;{h \over 2^n},0)$ (upper row) and $\mathcal{N}_{\bf \bar q}(t;
		{h \over 2^n},100)$ (bottom row), for $n=1,2,3$. The dashed horizontal line
		corresponds to first-order convergence $\mathcal{N}=1$. When the
		full grid is taken into account in the computation of the norm, we
		typically observe first-order convergence.
		On the other hand, by neglecting the last 100 grid points in the
		computation of the norm, we observe convergence of about
		$\mathcal{N} \simeq 1.5$ for $\Omega$ and $Y$ and
		$\mathcal{N} \simeq 2$ for $Z$.}
	\label{fig:stable_norm_convergence}
\end{figure*}

\begin{figure*} \centering
	\includegraphics[width=2.1\columnwidth,height=0.8\columnwidth]
	{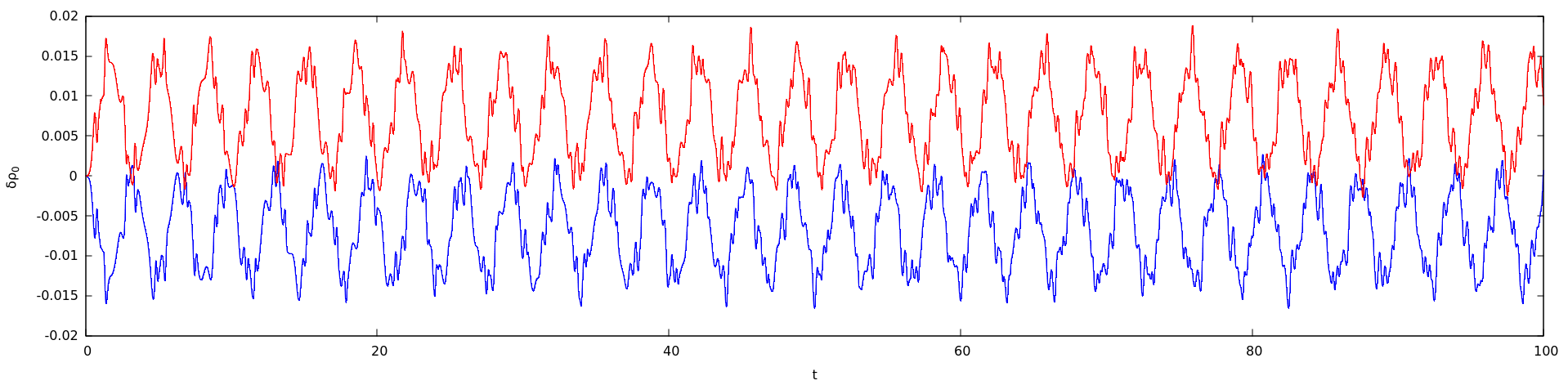}
	\caption{Stable stationary star: Central density perturbation against time
		for approximately $30$ oscillation periods. Red curve corresponds to the
		stationary initial data with $p_\omega = 0.001$ and blue curve with
		$p_\omega = -0.001$.}
	\label{fig:stable_rho_centre}
\end{figure*}

Let us first consider the stable stationary solution.
In Fig.~\ref{fig:stable_pointwise_convergence}, we plot $2^n \delta {\bf
\bar q}_{h \over 2^n}$ at four different resolutions $n=1,2,3,4$.
These are again plotted at three different times (rows), $t \simeq
0.04, 1.2, 4.2$.
We only show the case $p_\omega = -0.001$
as the case where $p_\omega = 0.001$ is qualitatively similar.
Note the different power of $h$ from
the dispersion/collapse case, due to the fact that we typically get
less than second-order convergence. The cause of this is an
instability originating from the outer boundary propagating
inwards. At the time $t \simeq 4.2$, this instability has moved to and
from the boundary twice. Equivalently, the time for the error
originating from the numerical outer boundary to reach the center is
$\Delta t \simeq 1.0$. As a consequence, the simulation losses its
second-order accuracy everywhere. There is also an instability at and
near the outer boundary that does not converge at all, but rather is
roughly equal at different resolutions. Nevertheless, as in the case
of dispersion, this instability propagates into the numerical grid
very slowly and its size shrinks with increased resolution.

In Fig.~\ref{fig:stable_norm_convergence}, we plot the convergence in the
norm. Due to the combination of the error originating from the outer
boundary and the error near the boundary not converging at all, we
find $\mathcal{N}_{\bf \bar q}(t;{h \over 2^n},0) \simeq 1$. Once the
region near the outer boundary is neglected by removing the last 100
grid points, we recover approximate second-order convergence
$\mathcal{N}_{\bf \bar q}(t;{h \over 2^n},100) \simeq 2$.

In Fig.~\ref{fig:stable_rho_centre}, we plot the oscillations in the
central density, $\delta \rho_0(t) := \rho_0(t)-\rho_0(0)$ for both signs
of the perturbation, $p_\omega = \pm 0.001$. The
simulation is run with $3200$ grid points, for sufficiently long time
so that the central density displays approximately $30$ cycles.
These oscillations maintain constant small amplitude, 
proportional to the initial perturbations, and we
conjecture that they are essentially linear oscillations with
constant frequency, as one would expect in a stable star.
The central density oscillates about an average that is offset from
the unperturbed star, because our perturbation of the initial data changes
the total mass of the star. Our unphysical copy outer boundary condition
does not seem to destroy this continuum property. Note that
when checking convergence, we only evolve the initial data up to at
most $t=4$. The reason is that for convergence testing, we consider
much higher resolution than we do in Fig.~\ref{fig:stable_rho_centre}.
Compare for example the highest resolution run ($n=5$, equivalent to
$25600$ gridpoints) when testing convergence, with the much lower
resolution used to produce Fig.~\ref{fig:stable_rho_centre} ($n=2$,
equivalent to $3200$ gridpoints).

\begin{figure*}
	\includegraphics[width=2.2\columnwidth,height=1.1\columnwidth]
	{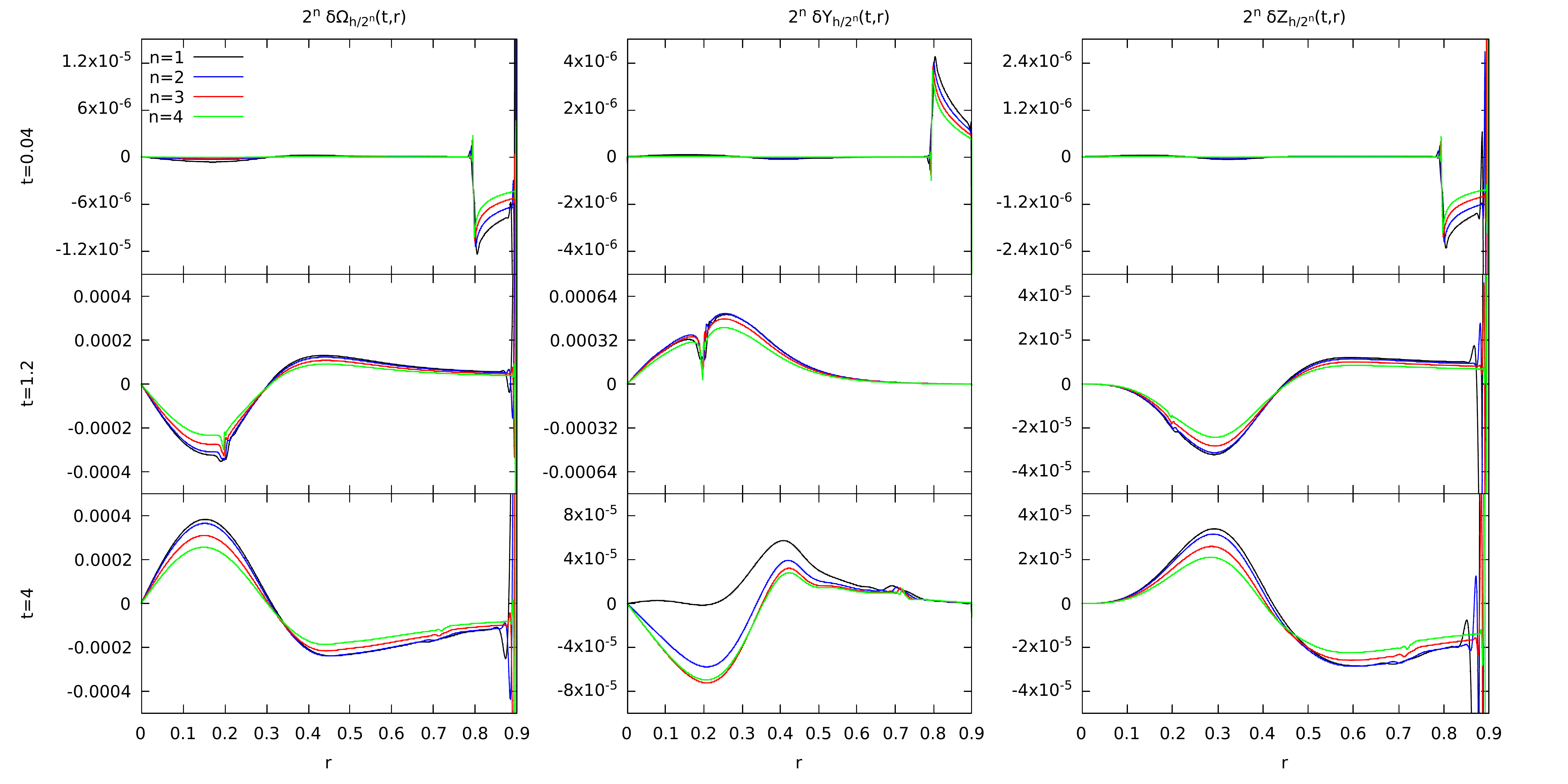}
	\caption{Unstable stationary star with negative density perturbation:
		A first-order error originating from the outer boundary travels inward,
		causing the evolution to converge only to first-order.
		Otherwise as in Fig.~\ref{fig:stable_pointwise_convergence}.}
	\label{fig:unstable_sub_pointwise_convergence}
\end{figure*}

\begin{figure*}
	\includegraphics[width=2.2\columnwidth,height=1.1\columnwidth]
	{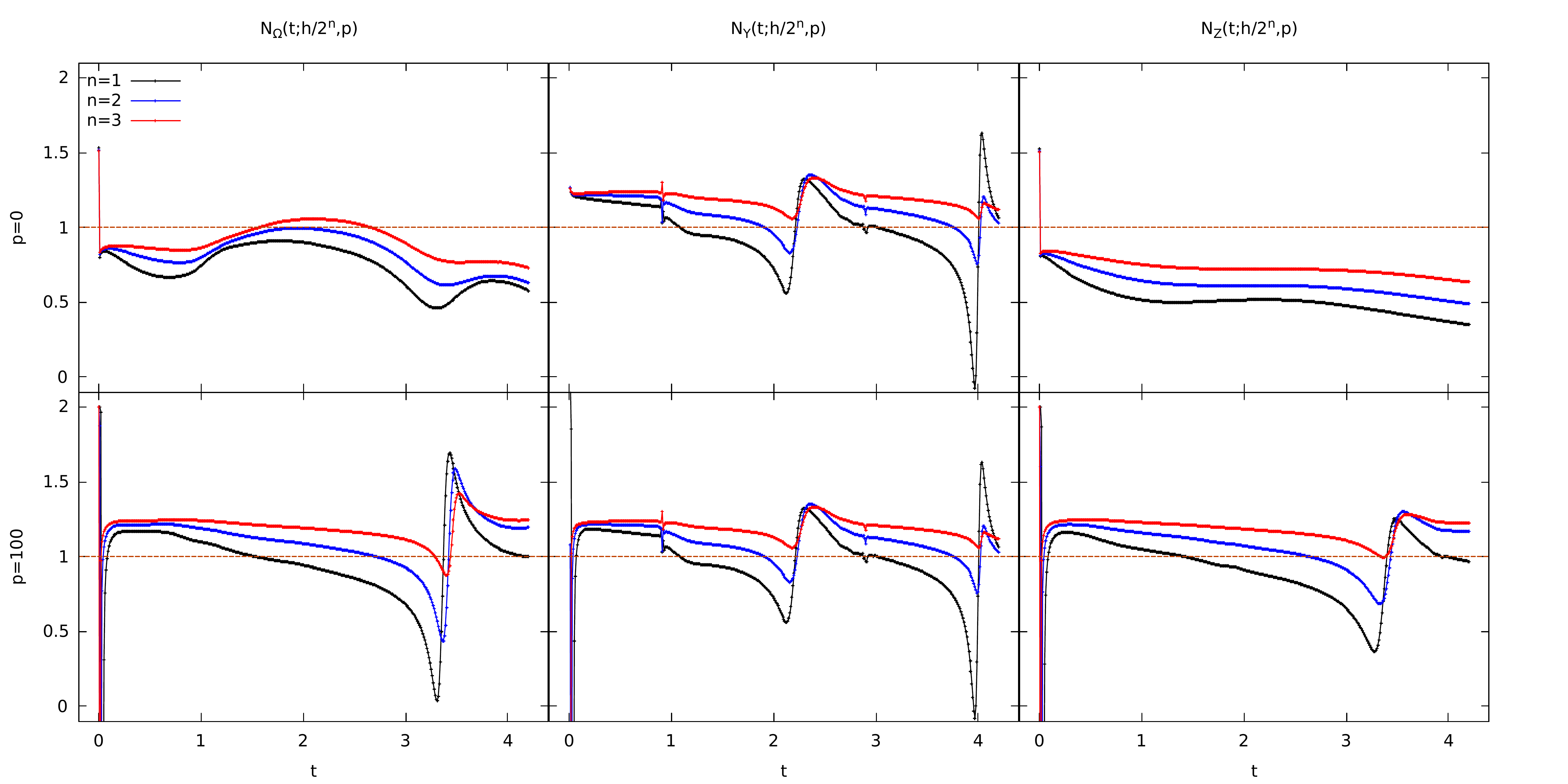}
	\caption{Unstable stationary star with negative density perturbation:
		Plots of $\mathcal{N}_{\bf \bar q}(t;{h \over 2^n},0)$ (upper row) and
		$\mathcal{N}_{\bf \bar q} ({h \over 2^n},100)$ (bottom row), for
		$n=1,2,3$. The dashed horizontal line corresponds to first-order
		convergence, $\mathcal{N}=1$. We find first-order
		convergence once the last 100 grid points are neglected.}
	\label{fig:unstable_sub_norm_convergence}
\end{figure*}

\begin{figure*}
	\includegraphics[width=2.1\columnwidth,height=0.8\columnwidth]
	{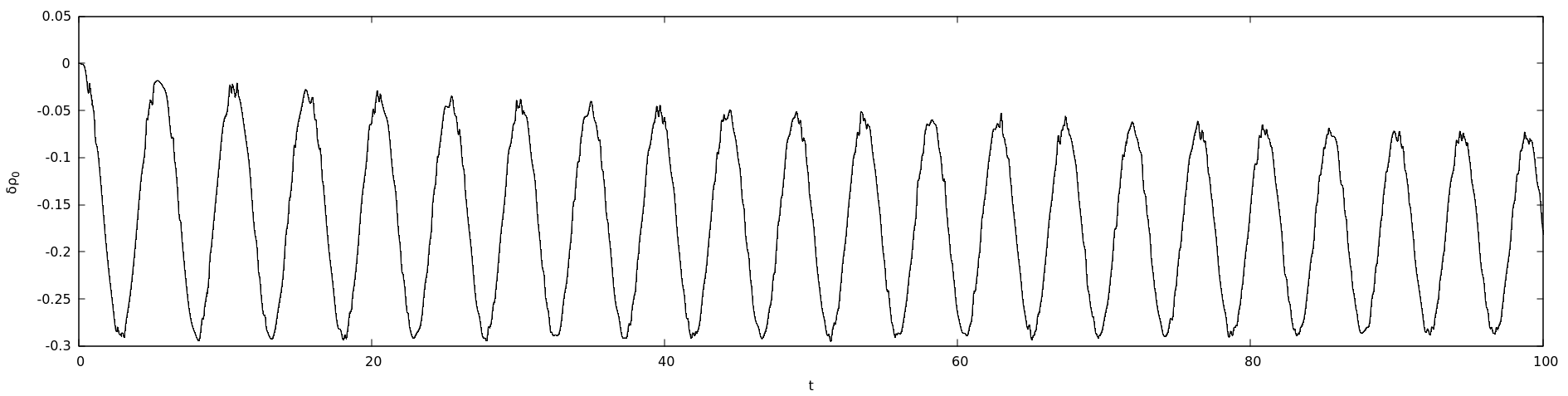}
	\caption{Unstable stationary star with negative density perturbation:
		Central density perturbation against time. The star breathes nonlinearly
		without collapsing. Compare with blue curve in
		Fig.~\ref{fig:stable_rho_centre}.}
	\label{fig:unstable_sub_rho_centre}
\end{figure*}

\begin{figure*}
	\includegraphics[width=2.2\columnwidth,height=1.1\columnwidth]
	{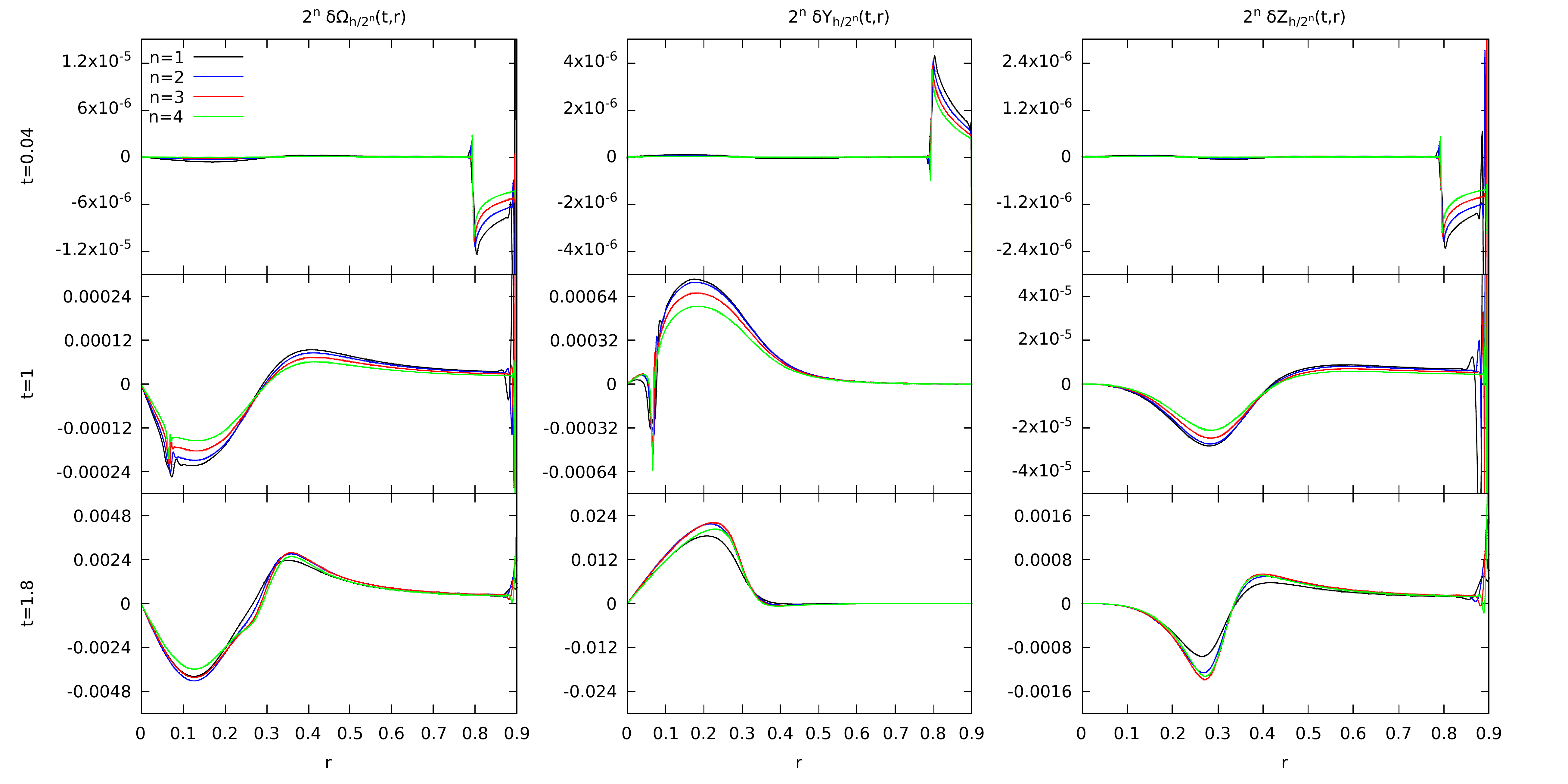}
	\caption{Unstable stationary star with positive density
		perturbation: We observe qualitatively similar behavior
		as for Fig.~\ref{fig:unstable_sub_pointwise_convergence}.}
	\label{fig:unstable_super_pointwise_convergence}
\end{figure*}

\begin{figure*}
	\includegraphics[width=2.2\columnwidth,height=1.1\columnwidth]
	{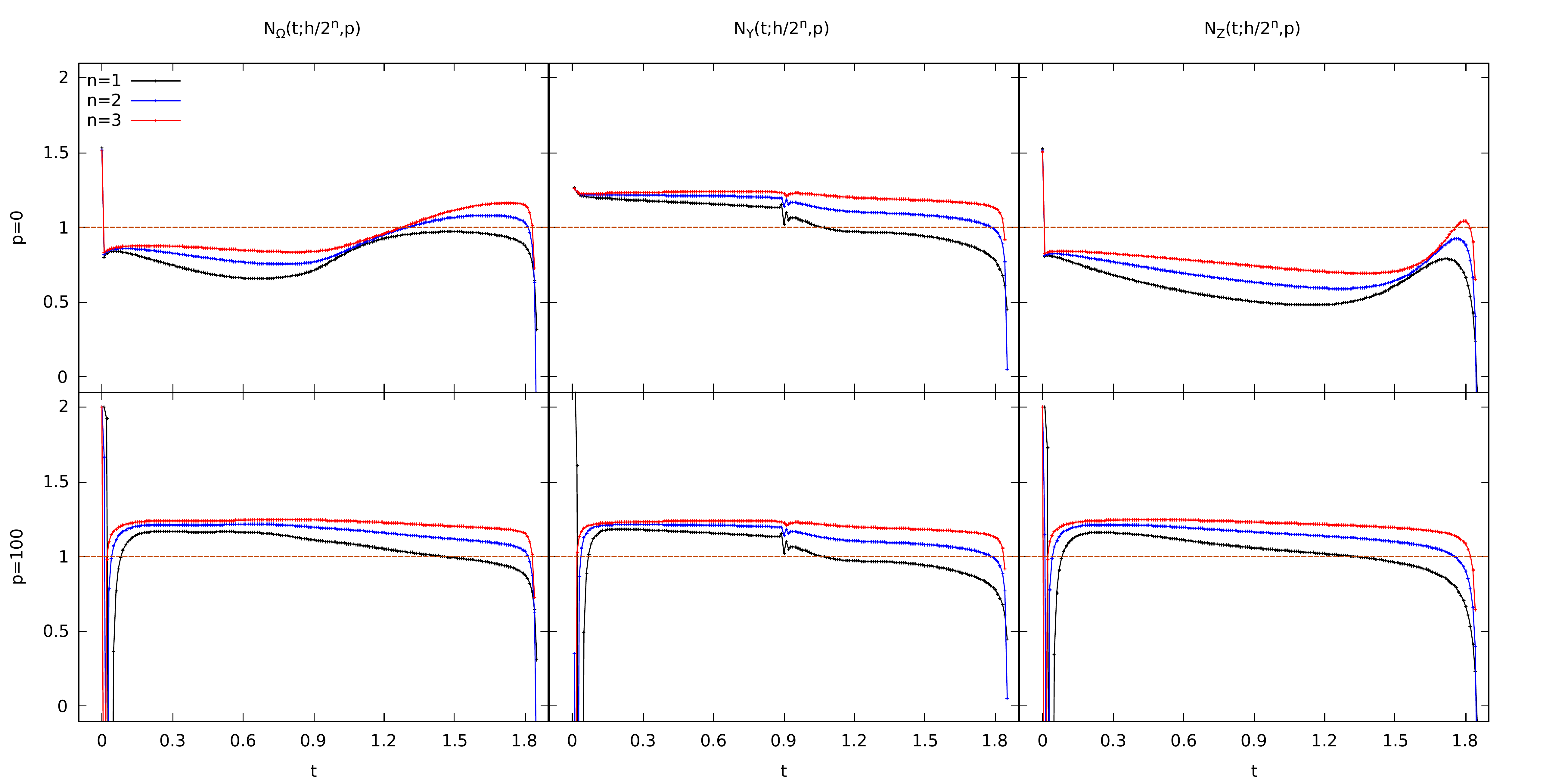}
	\caption{Unstable stationary star with positive density perturbation:
		Plots of $\mathcal{N}_{\bf \bar q}(t;{h \over 2^n},0)$ (upper row) and
		$\mathcal{N}_{\bf \bar q} (t;{h \over 2^n},100)$ (bottom row), for
		$n=1,2,3$. The dashed horizontal line corresponds to first-order
		convergence. We find here fairly constant convergence of 
		$\mathcal{N}_{\bf \bar q}(t;{h \over 2^n},100) \simeq 1.2$ up until
		the onset of collapse.}
	\label{fig:unstable_super_norm_convergence}
\end{figure*}

\begin{figure*}
	\includegraphics[width=2.2\columnwidth,height=1.1\columnwidth]
	{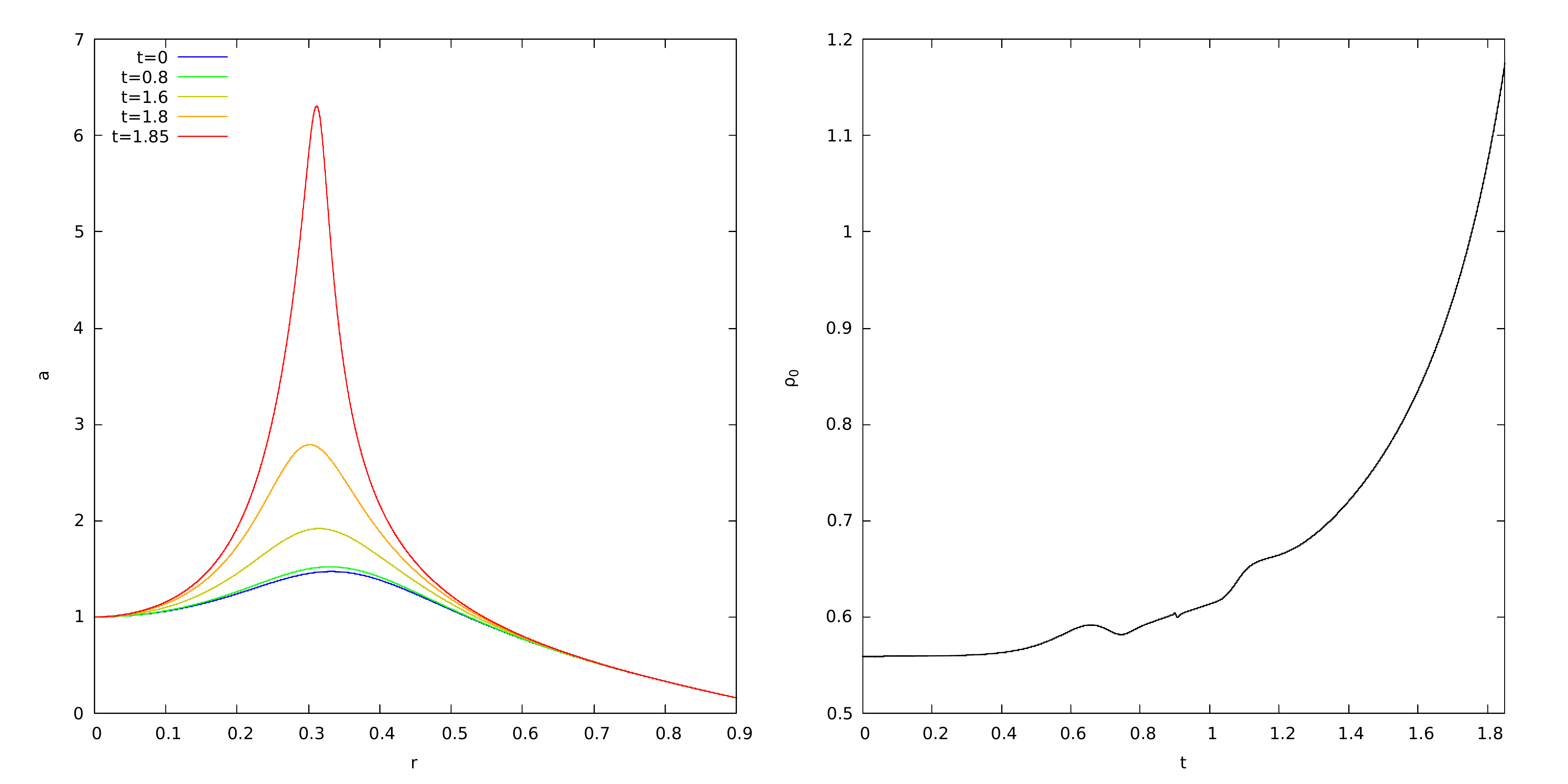}
	\caption{Unstable stationary star with positive density perturbation:
		We plot the metric coefficient $a(t_i,r)$ (left) at different times
		$t_i$, ranging from the initial time to the onset of collapse,
		as well as the central density against time (right). For this
		(positive) sign of the initial density perturbation, the
		star promptly collapses. Black-hole formation is triggered due
		to the timestep becoming small ($\Delta t \sim 10^{-10}$).}
	\label{fig:unstable_super_a_rho_centre}
\end{figure*}

Let us now turn to the unstable stationary solution.
The convergence tests for both cases are summarized in
Figs.~\ref{fig:unstable_sub_pointwise_convergence} and
\ref{fig:unstable_sub_norm_convergence} ($p_\omega=-0.001$) and
Figs.~\ref{fig:unstable_super_pointwise_convergence} and
\ref{fig:unstable_super_norm_convergence} ($p_\omega=0.001$).

Recall that for the unstable configuration, we have \textit{not}
imposed the positivity of the HLL flux for $Z$
at the outer boundary, as we heuristically find that otherwise a small
shock forms during the evolution, which prevents the simulation to
converge to the desired order in the norm. On the other hand,
lifting this constraint on the flux of $Z$
causes a first-order error originating from the outer boundary to
propagate inwards. The time for this error to reach the center
(for both signs of $p_\omega$) is $\Delta t \simeq 0.9$.
The simulation is only about first-order accurate.

As for the stable configuration, there is also an instability at and
near the outer boundary which does not converge at all, but rather is
roughly equal at different resolutions. Nevertheless, this instability
propagates into the numerical grid very slowly and its size shrinks
with increased resolution. Such a behavior can also be noted for the
stable configuration discussed above if the constraint on the
positivity of the HLL flux of $Z$ is removed there. In particular, a more
careful treatment of the boundary conditions at the numerical outer
boundary will be needed to accurately evolve the stationary solutions.

For a positive sign of the initial density perturbation, the
star promptly collapses into a black hole, see
Fig.~\ref{fig:unstable_super_a_rho_centre}. On the other hand, for a negative
sign, the star does not collapse. Instead, it breathes, i.e. the
central density oscillates periodically with very large
amplitude, down to about half of the stationary value. This can
be seen in Fig.~\ref{fig:unstable_sub_rho_centre}, where we plot the
central density perturbation $\delta \rho_0(t)$ at sufficiently long
times for $30$ cycles. The
simulation is run with $3200$ grid points as well. The local
maxima stay approximately constant throughout the simulation, and the
central density is approximately periodic.

It should be again emphasized that due to the fluctuating numerical
convergence for the stable and oscillating unstable cases (see again
Figs.~\ref{fig:stable_norm_convergence} and \ref{fig:unstable_sub_norm_convergence}),
it is uncertain how much of Fig.~\ref{fig:stable_rho_centre} and
Fig.~\ref{fig:unstable_sub_rho_centre} is physical or a numerical
effect. Nonetheless, we can already observe qualitative differences in
the evolution between the stable and unstable stationary initial data
even at short times.

\section{Conclusions}
\label{section:section6}

In this paper, we have presented a new code to simulate the
Einstein-fluid equations in axisymmetry in $2+1$ dimensions. We have
focused on the ultrarelativistic equation of state $p=\kappa
\rho$. However it should be straightforward to adapt the code to an
arbitrary barotropic or hot equation of state.

In the case of generic initial data that disperse or collapse both
with small and large angular momenta, we have demonstrated that the
code converges to second order in resolution both pointwise and
in the $\ell^2$ norm, except at and near the numerical
outer boundary, and near the onset of black hole collapse for highly
rotating configurations.

We have also evolved stable and unstable rotating stationary
stars. For these, the code converges only to first order.
Nevertheless, we can clearly distinguish stable and unstable stars,
even at short times. The former remain approximately stationary, with
only small oscillations, while the latter show two distinct evolutions
depending on the sign of the perturbation that we apply it to,
either collapse or very large (but still periodic)
oscillations. This provides some evidence in favor of our claim in
\cite{Carsten20}, where it was suggested that the family of stationary
stars with $|J| \leq M \ell$ is divided into two families of stable and
unstable solutions.

A fundamental strength of our approach is that we make full use of the
existence of two conserved matter currents (unexpectedly, for energy
as well as, expectedly, for angular momentum) and related local
expressions for the mass $M$ and angular momentum $J$. As a
consequence the metric evolution is fully constrained, and $M$ and
$J$ are exactly conserved.

A well-known disadvantage of polar-radial coordinates is that our code
stops as an apparent horizon is approached. However, one could in
principle make equal use of the two conserved currents and conserved
quantities in other coordinates. 

The main weakness of our code as presented here is that we have not
found a way of extending the outer boundary all the way to the
timelike infinity of any asymptotically BTZ spacetime, in a way that
is stable and accurate \cite{Patrick21}. This means that we have to impose
an unphysical ``copy'' boundary condition at finite radius
$R$. Fortunately, it turns out that, with some fine-tuning, this does
not prevent us from carrying out long-term (many sound-crossing times)
evolutions of stars. Moreover, it also does not seem to be an obstacle
in the investigation of critical phenomena at the threshold of
(prompt) collapse, which we will report on in a companion paper.

\begin{acknowledgements}
We are grateful to Ian Hawke for providing elements of our code.
Patrick Bourg was supported by an EPSRC Doctoral Training Grant to
the University of Southampton. 
\end{acknowledgements}


\end{document}